\documentclass[a4paper,superscriptaddress,twocolumn,prb,showkeys,preprintnumbers,floatfix,showpacs]{revtex4-2}

\usepackage{grffile}
\usepackage{simplewick}
\usepackage{epsfig}
\usepackage{amsmath}
\usepackage{amssymb}
\usepackage{blkarray, multirow, graphicx, diagbox, color, xcolor, colortbl}
\usepackage{ulem} 

\usepackage{braket}
\usepackage{tabularx}
\usepackage{booktabs} 
\newcolumntype{C}{>{$}c<{$}}
\AtBeginDocument{
	\heavyrulewidth=.08em
	\lightrulewidth=.05em
	\cmidrulewidth=.03em
	\belowrulesep=.65ex
	\belowbottomsep=0pt
	\aboverulesep=.4ex
	\abovetopsep=0pt
	\cmidrulesep=\doublerulesep
	\cmidrulekern=.5em
	\defaultaddspace=.5em
}

\usepackage{ifthen}
\usepackage{ulem}
\usepackage{tikz}
\usepackage{calc}
\usetikzlibrary{external}
\usepackage{pgffor}
\usepackage{pgfplots}
\pgfplotsset{compat=1.8}
\usepackage{pgfplotstable}
\usepgfplotslibrary{groupplots}

\definecolor{colorA}{rgb} {0.58,0,0.8275}
\definecolor{colorB}{rgb} {0.11,0.663,0.51}
\definecolor{colorC}{rgb} {0.3373,0.7059,0.9137}
\definecolor{colorD}{rgb} {0.902,0.6235,0}
\definecolor{colorE}{rgb} {0.9451,0.902,0.3255}

\usetikzlibrary
{
	calc,
	patterns,
	positioning,
	petri,
	arrows,
	decorations.markings,
	backgrounds,
	fit,
	graphs,
	shapes.geometric,
	decorations.pathmorphing,
	shapes.misc,
	shapes,
	tikzmark
}

\newcommand{\tJ}{$t$-$J$ }

\usepackage{xparse}

\newboolean{buildtikzpics}
\setboolean{buildtikzpics}{false} 

\newboolean{simpletrue}
\newboolean{simplefalse}
\setboolean{simpletrue}{true}
\setboolean{simplefalse}{false}

\usepackage{hyperref}

\usepackage{cleveref}

\begin{document}

\title{Lifshitz transition in the phase diagram of two-leg \tJ ladder systems at low filling}
\author{Steffen Bollmann}
\affiliation{Institut f\"{u}r Theoretische Physik, Georg-August-Universit\"{a}t G\"{o}ttingen, Friedrich-Hund-Platz 1, D-37077 G\"{o}ttingen, Germany}
\affiliation{Max Planck Institute for Solid State Research, 70569 Stuttgart, Germany}
\author{Alexander Osterkorn}
\affiliation{Institut f\"{u}r Theoretische Physik, Georg-August-Universit\"{a}t G\"{o}ttingen, Friedrich-Hund-Platz 1, D-37077 G\"{o}ttingen, Germany}
\author{Elio J. K\"onig}
\affiliation{Max Planck Institute for Solid State Research, 70569 Stuttgart, Germany}
\author{Salvatore R. Manmana}
\affiliation{Institut f\"{u}r Theoretische Physik, Georg-August-Universit\"{a}t G\"{o}ttingen, Friedrich-Hund-Platz 1, D-37077 G\"{o}ttingen, Germany}

\date{\today}

\begin{abstract} 
We use a combination of numerical matrix product states (MPS) and analytical approaches to investigate the phase diagram of the two-leg \tJ ladder in the region of low to intermediate fillings. 
We choose the same coupling strength along the leg- and rung-directions, but study the effect of adding a nearest-neighbor 
repulsion $V$.
We observe a rich phase diagram and analytically 
identify a Lifshitz-like
band filling transition, which 
can be associated to a numerically observed crossover from s-wave to d-wave like superconducting quasi-long range order (QLRO). 
Due to the strong interactions, the Lifshitz transition is smeared into a crossover region which separates two distinct Luttinger theories with unequal physical meaning of the Luttinger parameter. Our numerically exact MPS results spotlight deviations from standard Luttinger theory in this crossover region 
and is consistent with Luttinger theory sufficiently far away from the Lifshitz transition.
At very low fillings, studying the Friedel-like oscillations of the local density identifies a precursor region to a Wigner crystal at small values of the magnetic exchange interaction $J/t$.
We discuss analytically how tuning parameters at these fillings 
modifies the phase diagram, and find good agreement with MPS results.
\end{abstract}

\maketitle

\section{Introduction}

A particularly interesting system of strongly correlated electrons is the $t$-$J$ model, which was originally introduced as a simplification for the strong coupling limit of the Hubbard model, as already realized in the 1970s by Spałek et al.~\cite{SPATEK1977375,tJ1977} and later in the context of cuprate superconductivity~\cite{tJoriginal1,ZhangRice1988}.
It possesses a rich phase diagram, and is believed to be a basic model for the study of high-temperature superconductivity \cite{highTc_original} (see, e.g., \cite{Dagotto1994}). 
In the context of the latter, it has been subject to numerous studies, but despite these efforts, its lack of integrability \cite{caux_integrability} allows the phase diagram to be well known only for one-dimensional systems \cite{Moreno2011}.
More recently, the advent of ultracold polar molecules \cite{Silke_science,Silke_science2,Silke_Nature,reviewmolecules,review_molecules2,focus_ultracoldmolecules,molecules_review3,Molecules_Progress_Science2017} on optical lattices \cite{Bloch:2008p943} inspired a generalization of the original $t$-$J$-model with fully tunable interactions \cite{Gorshkov1,Gorshkov2}.
While its experimental realization is an ongoing challenge, progress has been made to demonstrate that spin exchange can be realized in these setups \cite{Yan:2013fn,PRLspindynamics_2014,preprint_itinerantspindynamics_molecules}, so that future investigations of the $t$-$J$ model in such experiments is envisaged.
On a chain, the phase diagram of variants of this model are well studied numerically, e.g., using matrix product states (MPS) \cite{Moreno2011,Schollwock:2011p2122}. 
The full tunability of interactions and the long-range nature of the dipolar interactions in the polar molecule setups show that interesting modifications can be achieved, e.g., an enhanced superconducting phase \cite{Gorshkov1,EPL_dipolartJchain,Manmana2017}, or topological SC \cite{PRL_tunabletJ_topoSC}.
However, the step towards exploring two-dimensional systems remains a major challenge.
Recent progress has been reported using iPEPS \cite{iPEPS_orig,PhysRevB.81.165104,MPSreviewVerstraete,review_tensorproductstates_CiracVerstraete,Corboz2DtJ,CorbozRVB,iPEPS2dtJ_vonDelft,Scipost_Grusdt_2020}, and also by treating multi-leg ladder systems (see, e.g., \cite{PRBKivelson_2017,PRB_tJFourlegladder_Rigol,PRBKivelson_2018,GrusdtPRB_2020,Bohrdt_2020,PNASWhite_2021,PRLSheng2021,white_preprint2022}).
In this context, the simplest non-trivial extension of the chain system is to treat two-leg ladder systems (see, e.g.,  \cite{PRBTroyer_1994,dagotto1996,PRBTroyer_1996,Hayward1996,PRBWhite_2001,WhiteAffleck2002,FeiguinAffleck2008,PRLWhite_2015,MusserSenthil2022,preprint_ladders}), which lately have been proposed to describe organic crystals, e.g., doped crystals of terphenyl~\cite{terphenyl_ladders}. 
The phase diagram has been investigated in some detail, but with an apparent focus on higher densities.
Here, we aim at complementing these studies by considering two directions of interest: i) investigate in more detail the behavior at low densities; ii) investigate the effect of tuning interaction strengths as previously studied in chain systems.
We pursue this goal by combining field theory, analytical considerations at very low fillings, and from a detailed numerical study using state-of-the art MPS methods. 
One important aspect is the 2-band nature of the two-leg ladder systems, leading to band-filling transitions \cite{Meng2011}. 
We show that such Lifshitz-type transitions are relevant for understanding the phase diagram at low fillings.
Notably, here it leads to a more complicated field theory interpolating between two known low-energy field theories.
In particular, in the two field theories the dependence of observables on the Luttinger parameter $K_c$ differs, so that in the crossover regime it is unknown how precisely the observables depend on $K_c$.

It is necessary to numerically compare in detail the behavior of observables such as the algebraic decay of correlation functions in order to map out the phase diagram.  
This is done in the following for two variants of the $t$-$J$ model, which allows us to study the effect of adding a Coulomb-type repulsion.
This has been found to enhance superconducting phases~\cite{Troyer1993,Manmana2017}, and here we can investigate the interplay with the Lifshitz-type transition. 
These considerations are complemented by an analytical treatment at very low densities, far below the Lifshitz transition, which allows us to estimate the extension of the phases when tuning parameters of the model, which can be useful in the context of future polar molecule experiments.

\begin{figure}
	\centering 
	\includegraphics[]{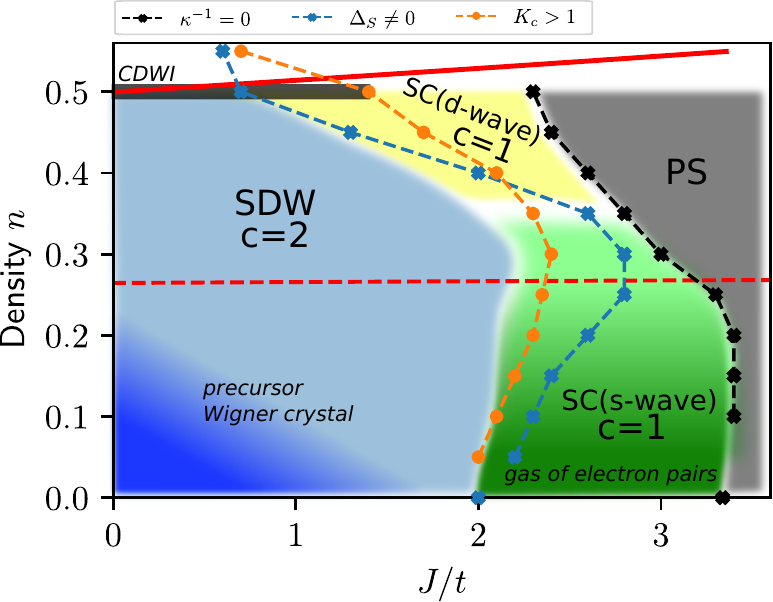}
	\includegraphics[]{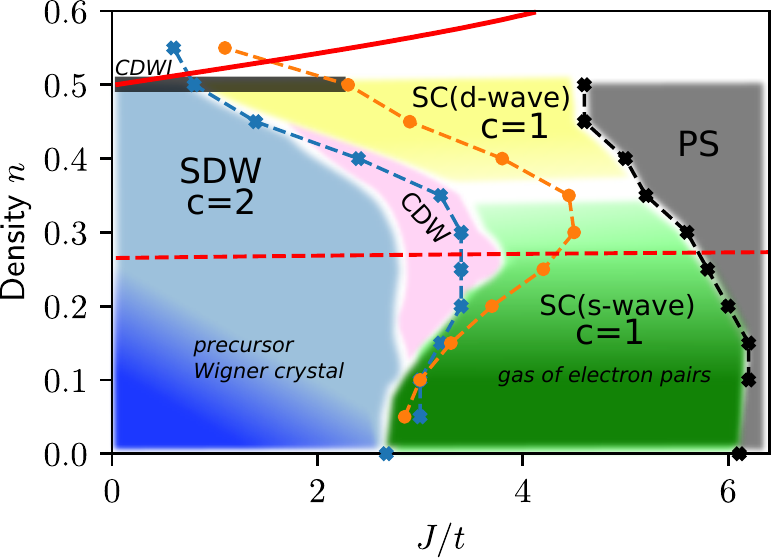}
	\caption{Ground state phase diagrams of (top) the regular and (bottom) the $V=0$ two-leg \tJ ladder \eqref{eq:H0} obtained using MPS and the Hartree-Fock approach to the band-filling transition of Sec.~\ref{sec:Lifshitz}. 
	SDW stands for a 2-channel Luttinger liquid (LL; with central charge $c=2$ as indicated) with dominant spin-density-wave correlations; CDW stands for a LL with dominant charge-density-wave correlations; SC stands for a $c=1$ LL with dominant pairing correlation functions. s-wave and d-wave indicate for the corresponding type of SC as discussed in Sec.~\ref{subsec:observables}.
	The precursor region to a Wigner crystal is identified via $4 k_F$-contributions to Friedel-like density oscillations. 
	PS stands for phase separation, identified by a diverging compressibility.
	At very low fillings, the s-wave SC phase is identical to a gas of free electron pairs. The bold red line denotes the occurrence of a Lifshitz transition according to the Hartree-Fock ansatz \eqref{eq:LifshitzCondition}. 
	The dashed red lines estimate the crossover region around the Lifshitz transition connecting the two different field theories at high and low densities, respectively. In this region, the Luttinger parameter $K_c$(orange circles indicate the line at which $K_c=1$ as obtained from the charge structure factor, see Secs. \ref{sec:AnalyticalExpectations}, \ref{sec:LuttParam}) cannot uniquely be determined using standard approaches. 
	All boundaries between the different LL phases are estimated by directly comparing the exponents of the different correlation functions. 
	The blue line denotes the opening of the spin gap. 
	The black horizontal line at $n=0.5$ indicates the opening of a charge gap inducing a symmetry broken CDW insulator (CDWI) phase.
	The two points at zero density are taken from the calculation in Sec.~\ref{sec:zero_density_limit}.}
	\label{fig:PhaseDiagramtJ}
\end{figure}

The paper is organized as follows. In Sec.~\ref{subsec:models}, we introduce the model and briefly discuss the two phase diagrams which are presented in Fig.~\ref{fig:PhaseDiagramtJ}. 
This is followed up in Sec.~\ref{subsec:analytics} by a presentation of analytical considerations regarding an appearing Lifshitz transition and the low energy field theory in means of bosonization.
In Sec.~\ref{subsec:observables}, we introduce the observables that we investigate to explore the ground state phase diagram. 
In Sec.~\ref{sec:Results}, we present details to how we obtain the phase diagrams from the DMRG data. 
In addition, in Sec.~\ref{sec:zero_density_limit} we discus how an analytical treatment of two electrons can be used to estimate the size of the superconducting phase in the zero density limit. 
We conclude with a summary in Sec.~\ref{sec:Summary}.
The appendices contain further details of the discussions and results presented in the main part of the paper.

\section{Our Setup: Models, Methods, Observables, and phase diagram}

\subsection{Variants of the $t$-$J$-Model on the Two-Leg Ladder Geometry}
\label{subsec:models}

In the following model, Eq.~\eqref{eq:H0}, we consider the usual $t$-$J$ model~\cite{SPATEK1977375, ZhangRice1988, auerbach2012, Dagotto1994}, but allow for a variable nearest-neighbor Coulomb repulsion $V$, which has been studied before (see, e.g.,  \cite{Troyer1993, Gorshkov1,Manmana2017}). 
On the two-leg ladder systems treated by us it reads 
\begin{equation}
\label{eq:H0}
 \mathcal{H}^{tJV}_{\text{ladder}} = \mathcal{H}^t + \mathcal{H}^{JV}_{\text{leg}} + \mathcal{H}^{JV}_\text{rung}
\end{equation}
with
\begin{align*}
	\mathcal{H}^t = & -t \sum_{i, l, \sigma}P_s\left[c^\dagger_{i, l, \sigma}c^{\phantom{\dagger}}_{i+1, l, \sigma} + \text{H.c.}\right]P_s \notag \\ & -t\sum_{i, \sigma} P_s\left[c^{\dagger}_{i, 1, \sigma}c^{\phantom{\dagger}}_{i, 2, \sigma} + \text{H.c.}\right]P_s,
\\
	\mathcal{H}^{JV}_{\text{leg}} = &J \sum_{i, l}  \left[\vec{S}_{i, l} \vec{S}_{i+1, l} - \frac{V}{4}n_{i, l}n_{i+1, l}\right],\\
	\mathcal{H}^{JV}_{\text{rung}} = &J \sum_{i}  \left[\vec{S}_{i, 1} \cdot \vec{S}_{i, 2} - \frac{V}{4}n_{i, 1}n_{i, 2}\right].
	\label{eq:model}
\end{align*}

Here, $l = 1,2$ is the leg index, $i$ labels the rung, $\sigma$ the spin direction, $P_s$ projects out double occupancies, 
$c_{i,l, \sigma}$ is a fermionic annihilation operator acting on site $(i,l)$ and spin-direction $\sigma$, $n_{i,l} = \sum_{\sigma} c^\dagger_{i,l,\sigma}c^{\phantom{\dagger}}_{i,l,\sigma}$ is the occupation operator, and $\vec S_{i,l} = c^\dagger _{i,l,\sigma} \vec{\sigma}_{\sigma \sigma'}^{\phantom{\dagger}} c^{\phantom{\dagger}}_{i,l,\sigma'}/2$. 
We set the lattice constant 
equal to unity and work in units in which $\hbar \equiv 1$.

In the following, we treat antiferromagnetic spin exchange $J>0$. For $V = 1$, we obtain the usual $t$-$J$ model as obtained from 2nd order perturbation theory in the strong coupling limit of the Hubbard model. 

In the following we will call this case the \textit{regular} \tJ ladder. 
Furthermore, we treat the system with $V=0$, which is obtained by adding the corresponding nearest-neighbor Coulomb-repulsion to the original \tJ model. This case will be referred to as the $V=0$ \tJ ladder.

One important effect of adding a Coulomb-repulsion is that it suppresses phase separation, and can lead to enhanced superconducting phases \cite{Troyer1993}. 
Here, we revisit its effect in the low-density regime of the two-leg ladder system.

In Fig.~\ref{fig:PhaseDiagramtJ} we present our main results and show the phase diagrams of both variants as obtained using MPS and the analytical approaches discussed further below.
As can be seen, the phase diagrams are quite similar to each other: both show a sequence from a gapless Luttinger-liquid (LL) phase with central charge $c=2$ and dominant spin correlations (SDW) at small values of $J/t$ to a $c=1$ LL with finite spin gap and dominant pairing correlation functions.
At even larger values of $J/t$, phase separation sets in.
These features and sequences of phases are very similar to the findings in \tJ chains, in particular at very low densities.
At the density $n=0.5$, a charge gap opens for small $J/t$ and a charge density wave insulator (CDWI) is formed.
The $V=0$ \tJ ladder has an enhanced SC phase, which has been observed also for the corresponding chain system \cite{Manmana2017} (note that throughout the manuscript we denote by SC a phase with dominant pairing correlation functions, i.e., a phase with SC quasi-long-range order).
In particular, the size of the spin-gap region before phase separation is substantially increased. However, in contrast to the chain, where the sequence of phases is not altered for $V=0$, here at a density around $n=0.3$ an additional $c=2$ LL phase with dominant CDW-correlations is realized. 
This raises the question, if and how changing $V$ in further \tJ-systems (e.g. broader ladder systems or in 2D) can lead to new features in the phase diagram.
Both systems show at low densities and small values of $J/t$ a precursor-region to a Wigner crystal, which is similar to the findings in Hubbard chains reported in Ref.~\onlinecite{Eggert2009}. 
In the spin-gap region, the s-wave SC at very low densities is identified to be a gas of electron pairs, which was previously reported also for the chain systems \cite{Moreno2011}.   

In the following, we explain in some detail how these phase diagrams were obtained and describe the field theoretical treatment around the Lifshitz transition line, which we estimate using a simple Hartree-Fock ansatz.

\subsection{Analytical considerations: Lifshitz transition}
\label{subsec:analytics}

In this section we summarize the analytical expectations for our system, Eq.~\eqref{eq:H0}. We consider a Hubbard-Heisenberg ladder instead of the constrained \tJ ladder, i.e. we consider

\begin{subequations}\label{eq:HU}
\begin{equation}
 \mathcal{H}^{UJV}_{\text{ladder}} = \mathcal{H}^t_0 + \mathcal{H}^U + \mathcal{H}^{JV}_{\text{leg}} + \mathcal{H}^{JV}_\text{rung}
 \label{eq:modelUsedInCalculations}
\end{equation}

with

\begin{align}
	\mathcal{H}^t_0 = & -t \sum_{i, l, \sigma}\left[c^\dagger_{i, l, \sigma}c^{\phantom{\dagger}}_{i+1, l, \sigma} + \text{H.c.}\right]  \\ & -t\sum_{i, \sigma} \left[c^{\dagger}_{i, 1, \sigma}c^{\phantom{\dagger}}_{i, 2, \sigma} + \text{H.c.}\right],
\\
\mathcal {H}^U & = U/2 \sum_{i, l} n_{i,l} (n_{i,l} - 1), \label{eq:HubbardU}
\end{align}

\end{subequations}

and all other terms as in Eq.~\eqref{eq:H0}. 
The Hubbard-Heisenberg ladder is equivalent to the t-J ladder only in the limit $U \rightarrow \infty$. Following a similar strategy to the one successfully applied in the literature \cite{WhiteAffleck2002} for high densities, we here study the Hubbard-Heisenberg ladder at small and intermediate $U/t$ using perturbation theory and bosonization. Clearly, there is no guarantee that our results would reproduce the $U \rightarrow \infty$ limit of the Hubbard-Heisenberg model, and of course perturbation theory and Gutzwiller projection do not commute. 

\subsubsection{Lifshitz transition at the Hartree-Fock-level}
\label{sec:Lifshitz}

The non-interacting tight binding Hamiltonian on a ladder with nearest neighbor hopping can be solved exactly by the basis transformation 
\begin{equation} \label{eq:0PiExpansion}
	c_{i, 0/\pi, \sigma} = \frac{1}{\sqrt{2}}\left(c_{i, 1, \sigma} \pm c_{i, 2, \sigma}\right)
\end{equation}
which decouples the system into two independent chains and brings the Hamiltonian into the form 
\begin{equation}
\mathcal{H}^{t} = -t\sum_{i, a, \sigma} \left[c^{\dagger}_{i, a, \sigma}c^{\phantom{\dagger}}_{i+1, a, \sigma}+\text{H.c.}\right] - t \sum_{i} (n_{i, 0} - n_{i, \pi}),
\end{equation}
where $a = 0,\pi$.
Each chain can independently be diagonalized 
resulting in dispersion relations
\begin{equation}
\varepsilon_{0/\pi}(k) = -2t\cos(k)\mp t.
\end{equation}
Thus, a Lifshitz transition occurs when the Fermi energy hits $E_F = - t$, which corresponds to an electron density of $n = \langle n_i \rangle = 1/2$ per site (quarter filling). 
Hence, based on the non-interacting estimate, there is a Lifshitz transition at the upper edge of the phase diagram presented in Fig.~\ref{fig:PhaseDiagramtJ}.

We now discuss interaction induced shifts of the position of this transition as displayed in Fig.~\ref{fig:PhaseDiagramtJ}.
Using a Hartree-Fock calculation (details are relegated to Appendix~\ref{app:HartreeFock}) we obtain an effective quadratic Hamiltonian
\begin{equation}
\mathcal{H}^{t} = -t_\Vert^*\sum_{i, a, \sigma} \left[c^{\dagger}_{i, a, \sigma}c^{\phantom{\dagger}}_{i+1, a, \sigma}+\text{H.c.}\right] - t_\perp^* \sum_{i} (n_{i, 0} - n_{i, \pi}).
\end{equation}
Under the assumption that only the lower band is occupied we find
\begin{subequations}
\begin{align}
  t_\Vert^* &= t+ J \frac{\sin(\pi n)}{8\pi} \left (3 -{V} \right) \label{eq:tPararenorm}\\
    t_\perp^* &= t + J \frac{n}{8} \left (3 - V \right) \label{eq:tperpRenorm}
\end{align}
\label{eq:Renormts}
\end{subequations}
and the dispersion
\begin{equation}
\varepsilon_{0/\pi}(k) = -2t_\Vert^*\cos(k)\mp t_\perp^*.
\end{equation}
Note that the Hubbard $U$ does not enter $t_\Vert^*$ or $t_\perp^*$, but merely leads to an overall energy shift in the spectrum. This fact is used as we push the Hartree-Fock calculation beyond the limits of its validity when we send $U \rightarrow \infty$ to enforce the equivalence between Hubbard-Heisenberg and \tJ models. 
By assumption of filling only the lower band we further employ $k_F = \pi n$ for the Fermi energy so that the Lifshitz transition is defined by the condition
\begin{equation} \label{eq:LifshitzCondition}
    -2t_\Vert^*\cos(\pi n) - t_\perp^* = -2t_\Vert^*+ t_\perp^*.
\end{equation}
This condition is displayed as a solid red curve in Fig.~\ref{fig:PhaseDiagramtJ}. Effectively, the spin-interaction increases the splitting by $t_\perp^*$ of the bands.

\subsubsection{Lifshitz transition beyond Hartree-Fock}

\begin{figure}
	\centering
	\includegraphics[scale = 1]{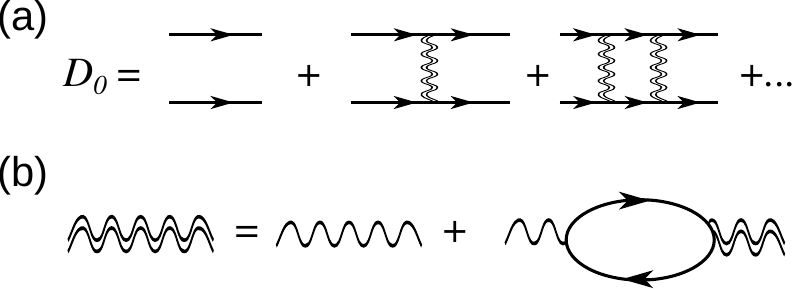}
	\caption{Two-particle Green's function in the upper subband. a) Diagrammatic representation of Eq.~\eqref{eq:LadderResummationLifshitz}. b) Diagrammatic representation of Eq.~\eqref{eq:RPAScreening}}
	\label{fig:Diagrams}
\end{figure}

Lifshitz transitions in interacting one-dimensional systems are non-trivially affected by the interactions in the system~\cite{Meng2011, Takahashi1971, Ogata1990, Goehmann1998, Essler2005}. We here briefly review this physics from different standpoints: First, when the chemical potential is far below the upper band, the effect of interband interactions only leads to virtual processes renormalizing the Luttinger liquid (LL) in the lower band, which can be treated using perturbation theory. There is a typical energy scale $E_p$ where this perturbation theory breaks down, i.e. the single band LL physics is inapplicable when the chemical potential is closer than $E_p$ to the upper band. As we explain in more detail below, this energy scale can also be estimated from the immediate vicinity of the Lifshitz transition. Second, when the chemical potential is far in the two-band regime, one may study the multi-band system using perturbative RG. Again, these RG equations break down inside a window of size $E_p$ above the bottom of the upper band. Third, we now discuss the physics inside this window, concentrating on a chemical potential $\mu_\pi \rightarrow 0^-$ right below the Lifshitz transition. Then, the free two-particle propagator in the upper subband takes the form $\mathcal D_\pi^{(0)}(k, \omega) = - i \sqrt{m/4\epsilon^+}$, with $\epsilon^+ = \omega-k^2/4m + 2 \mu_0 + i 0$ (in our case, $2m =1/t_\Vert^*$). In the presence of an intrasubband interaction $\mathcal {V}$ within the upper band  (but for the moment neglecting interband interactions), particles repeatedly scatter off each other resulting in an exact inverse two-particle Green's function, Fig.~\ref{fig:Diagrams}  a),

\begin{equation}\label{eq:LadderResummationLifshitz}
    [D_\pi(k, \omega) ]^{-1}= [D_\pi^{(0)}(k, \omega)]^{-1} -\mathcal{V}.
\end{equation}
Clearly, the second term always dominates for on shell excitations $\omega = k^2/4m$ as soon as $\vert \mu_\pi \vert < E_p \equiv m\mathcal{V}^2/8$. 
Differently said, the particles at the Lifshitz transition are always strongly interacting and form a so-called ``impenetrable electron gas'', in which the two-electron wave function has nodes at equal particle positions even for opposite spin.
As a consequence, when the lower band is coupled to this impenetrable gas, pair tunneling of singlet Cooper pairs is 
strongly suppressed and it was shown that also spin-spin interband interactions are marginally irrelevant. It was found in Ref.~\cite{Meng2011} that the lower subband remains a LL and ultimately screens the single particle excitations in the upper subband thereby suppressing their quasi particle weight. It is thus reasonable to think of a Lifshitz transition of polarons instead of electrons. 

The interaction $\mathcal V$ is the most relevant perturbation at the critical point~\cite{Meng2011} and defines the crossover scale $E_p$. 
We estimate this coupling for the model~\eqref{eq:HU}, keeping in mind that we want to ultimately push our theory to the $U \rightarrow \infty$ limit.
To leading order, $\mathcal V \sim U$ 
and we dropped weaker intersite interactions. The interaction constant $\mathcal V$ is screened and for $\mu_\pi \rightarrow 0^-$ only the filled subband can contribute to static screening. Using the Dzyaloshinski-Larkin theorem\cite{DzyaloshinskiiLarkin1973, Giamarchi2003}, according to which the random-phase-approximation (RPA) is exact for a linearized spectrum in 1D, we obtain the following effective interaction, Fig.~\ref{fig:Diagrams} b),
\begin{equation}\label{eq:RPAScreening}
    \mathcal V_{\rm eff}(\omega, k)^{-1} =\mathcal V^{-1} + \Pi(\omega,k),
\end{equation}
with $\Pi(\omega,k) = v_{F,0} k^2/\pi [(\omega + i0)^2 - v_{F,0}^2k^2]$ and $v_{F,0} = 2 t^*_\Vert\sin(\pi n)$. This interaction enters the ladder resummation, Eq.~\eqref{eq:LadderResummationLifshitz}, and we exploit that for on-shell two-particle excitations in the upper band the static screening approximation is justified, i.e. $\Pi(\omega = k^2/4m,k) \simeq 1/\pi v_{F,0}$. Physically, this follows from the much faster dynamics in the lower as compared to the upper subband. In summary we find $\mathcal V \simeq \pi v_{F,0}$ and $E_p = \pi^2 t_\Vert^* \sin^2(\pi n)/2$.

We also comment on the regime of partially, but dilute filling of the upper subband. It is theoretically harder to describe, yet the same crossover scale $E_p$ is believed to limit the strongly interacting regime for positive $\mu_\pi$, as well\cite{Meng2011}. Approaching $\mu_\pi \rightarrow E_p$ from above, we estimate the screening of the bare interaction by Eq.~\eqref{eq:RPAScreening} and $\Pi(\omega,k) \approx 1/\pi v_{F,0} + 1/\pi v_{F,\pi}$ which is determined by the Fermi velocities of both filled bands.

Using $E_p = m \mathcal V^2/8$ both above and below the Lifshitz transition, the window of impeneatrable electron gas is bounded by
\begin{equation} \label{eq:LifshitzWindo}
    -2t_\Vert^*\cos(\pi n) - t_\perp^* = -2t_\Vert^*+ t_\perp^* \pm E_p.
\end{equation}

These conditions are displayed as dashed red lines in Fig.~\ref{fig:PhaseDiagramtJ}. We remark in passing that above the Lifshitz transition $\pi n = k_F^\pi + k_F^0$, we will use this relationship in the remainder of the paper.

\subsubsection{Low-energy field theory}
\label{sec:fieldtheory}

Interacting one-dimensional fermionic systems are field theoretically suitably captured by means of bosonization~\cite{Giamarchi2003} in the limit when the important energy scales are small with respect to the Fermi energy counted from the edges of a given band ($x_i = \alpha i$, where we have introduced the lattice constant $\alpha$ for clarity), i.e.,
\begin{align}
    c_{i,a, \sigma} \sim \sum_\pm e^{\pm i k_F^a x_i - i (\pm \Phi_{a,\sigma}(x_i) - \Theta_{a,\sigma}(x_i))}.
\end{align}

Here, $\Phi_{a,\sigma}(x), \Theta_{a,\sigma}(x)$, are conjugate fields which are slow on the scale of the lattice constant and $k_F^a$ is the Fermi momentum in band $a = 0, \pi$. It is convenient to introduce bosonic fields in the charge and the spin channels $c$ and $s$, respectively,
\begin{equation}
    \Phi_{a,c/s} = \frac{\Phi_{a,\uparrow} \pm \Phi_{a, \downarrow}}{\sqrt{2}}
\end{equation}
so that, in bosonic language, the kinetic part of the action takes the form
\begin{equation} \label{eq:Skin}
    S_{\rm kin} = \sum_{\substack{\zeta = c,s\\a = 0,\pi}} \frac{1}{2\pi K_{\zeta,a}} \int d\tau dx\; \left [\frac{\dot \Phi^2_{a,\zeta}}{v_{\zeta, a}} + v_{\zeta,a} {\Phi'_{a,\zeta}}^2 \right].
\end{equation}
Here, $K_{\zeta,a}$ is the Luttinger parameter and $v_{\zeta,a}$ the Luttinger velocity, respectively, of the corresponding degree of freedom.  
In the non-interacting limit $K_{\zeta,a} = 1$ and, of course, the bosonic fields in the $a = \pi$ subband only exist above the Lifshitz transition.

In the presence of interactions, the situation is more subtle as some of the bosonic degrees of freedom gap out. It is custumary to summarize these interacting phases by the label C$n$S$m$ where $n$ and $m$ denote the number of gapless bosonic modes in the charge and spin channel, respectively~\cite{LinBalentsFisher1997}. A means to efficiently extract the total number of gapless modes numerically is to measure the central charge $c = n + m$.
Numerically, this can be done by analyzing the spatial behavior of the von Neumann entanglement entropy, which for systems with open boundary conditions is given by \cite{HolzheyWilczek1994,Korepin2004,Calabrese04}
\begin{equation}
    \mathcal S(x) = \frac{c}{6} \ln \left [\frac{L}{\pi} \sin\left (\frac{\pi x}{L}\right) \right] + d,
    \label{eq:vonNeumannEntropy}
\end{equation}
where $x$ is the position of the bipartition and $d$ a nonuniversal constant. 
$\mathcal{S}(x)$ is easily computed by MPS \cite{Schollwock:2011p2122}. 
However, the open boundary conditions (OBC) lead to additional oscillations in $\mathcal{S}(x)$, which can be understood in terms of the oscillations of the local kinetic energy \cite{PRL_Laflorencie_2006}. 
These are of the form \cite{Roux2008}
$B(t(l, i)- \Bar{t})$, where $B$ is a free constant that needs to be fitted, $t(l,i)=\sum_{\braket{(l, i),(h, j)}}\braket{c^\dagger_{(h,j)}c^{\phantom{\dagger}}_{(l,i)}}$ is the local kinetic energy and $\Bar{t}$ is the mean value of $t(l,i)$ in the bulk. The resulting central charge 
is indicated in the phase diagrams of  Fig.~\ref{fig:PhaseDiagramtJ}; more details of our procedure can be found in Appendix~\ref{app:entropy}.

We now review the analytical expectations, and first consider $\mu_\pi < - E_p$. In this limit, the $a = \pi$ band may be disregarded and the effective low-energy theory is a spinful interacting single band model in the band of $a = 0$ orbitals. The charge mode remains gapless with nonuniversal $K_{c,0}$. The spin mode is gapless with $K_{0,s} = 1$ for small $J/t$ (a C1S1 phase with $c = 2$) but gaps out for sufficiently large spin interactions leading to a C1S0 phase. Technically this is defined by the condition that the interaction constant mixing chiral and antichiral spin currents changes sign \cite{Giamarchi2003, Gogolin2004}.

Next, consider the regime $\mu_\pi >E_p$. In this limit, the bosonization of the two-leg Hubbard model is justified for moderate $U$ and weak coupling. Renormalization group calculations\cite{LinBalentsFisher1997, WhiteAffleck2002} predict that out of the four bosonic modes, only the field describing the total charge,
\begin{equation}
    \Phi_{+,c} = \frac{\Phi_{0,c}+\Phi_{\pi,c}}{\sqrt{2}},
\end{equation}
remains gapless. This corresponds to a C1S0 phase with action
\begin{equation}
    S_{\rm C1S0}  = \frac{1}{2\pi K_{c,+}} \int d\tau dx \; \left [\frac{\dot \Phi^2_{+,c}}{v_{+,c}} + v_{+,c} {\Phi'_{+,c}}^2 \right],\label{eq:C1S0}
\end{equation}
with $K_{c,+}$ and $v_{+,c}$ the corresponding LL parameter and velocity. 

Finally, in the regime $-E_p < \mu_\pi < E_p$ of the impenetrable electron gas around the Lifshitz transition, a direct bosonization approach is challenging. Indeed, in the regime $0< \mu_\pi <E_p$ the interaction energies are large as compared to the chemical potential $\mu_\pi$ of the upper subband, while in the regime $-E_p<\mu_0<\pi$, the perturbative inclusion of virtual interband processes breaks down. Yet, one may hope that qualitative aspects of the physics above/below the Lifshitz transition persist in the impenetrable crossover regime and this work constitutes a numerical exploration thereof. 
Indeed, our numerical results indicate that the central charge continues to be either 1 or 2, so that also in this intermediate regime the phases can be characterized as C$n$S$m$ phases. However, this regime realizes a crossover region of two different effective field theories, Eqs.~\eqref{eq:Skin} and~\eqref{eq:C1S0}, respectively.
Usually, extracting the LL parameter from the numerics is possible and the results can be used to characterize the different phases.
However, in such a crossover region of two different low-energy field theories, this approach is not as straight-forward in the following way:
as discussed further below, the structure factor follows the expectation from bosonization (in particular a linear behavior at small momenta $k$), so that the usual procedure allows one to extract a numerical value for the LL parameter in the charge sector.
This is the value plotted in the phase diagrams of Fig.~\ref{fig:PhaseDiagramtJ}.
However, following Tab.~\ref{tab:Expectations}, the linear coefficient of the structure factor has to interpolate between
$K_{c,0}$ and $K_{c,+}$ of the field theories Eqs.~\eqref{eq:Skin} and~\eqref{eq:C1S0} with an unknown function. 

How does this affect observables? Schematically, the electronic operators are $c_{i,\sigma} \sim \sum_\pm e^{i[\mp  \Phi_{0,c} + \Theta_{0,c}]/\sqrt{2}}$ below the Lifsitz transition, but $c_{i,\sigma} \sim \sum_\pm e^{i[\mp \Phi_{+,c} + \Theta_{+,c}]/2}$ above the transition. 
Thus, exponents stemming from $\langle \Phi_c(x) \Phi_c(0) \rangle$ correlators, which are $\sim 1/K_{c,0}$ below the transition become  $\sim 1/2K_{c,+}$ above the transition, while those exponents stemming from $\langle \Theta_c(x) \Theta_c(0) \rangle$ correlators change from $\sim K_{c,0}$ to $K_{c,+}/2$. The latter correlators however do not show up in the observables we study, see below.

We remark in passing that the commensurate filling at density $n =1/2$ is special as it allows for a charge density wave state with gapped charge sector (charge density wave insulator, CDWI).

\subsection{Observables and analytical expectations}
\label{subsec:observables}

In this section, we introduce the calculated observables we used to determine the phase diagrams, based on the approach of Refs.~\onlinecite{Moreno2011,Manmana2017}. 

\subsubsection{Global observables}

We define the spin gap as 
\begin{equation}
    \Delta_S = E_0(N,S^z_\text{total}=1) - E_0(N,S^z_\text{total}=0) \, ,
\end{equation}
where $E_0$ is the groundstate energy of a finite system with $L$ lattice sites and $N$ particles in the corresponding spin sector $S^z_\text{total}$. 
We extrapolate to the thermodynamic limit (TL) by keeping the density $n=N/L$ fixed and taking $L \to \infty$.   

In a similar fashion also a charge gap can be defined via 
\begin{equation*}
    \Delta_C = \mu_+ - \mu_-,
\end{equation*}
where $\mu_+$ and $\mu_-$ are the energies needed to add or remove a particle, respectively. However, since adding or removing one particle would necessarily change $S^z_\text{total}$ by one, this quantity can be influenced by the behavior of the spin gap. Therefore, we define the charge gap by adding and removing two particles to keep $S^z_\text{total}$=0, i.e., 
\begin{align}
    \Delta_C =& E_0(N+2,S^z_\text{total}=0) + E_0(N-2,S^z_\text{total}=0)  \nonumber \\ &-E_0(N,S^z_\text{total}=0).
\end{align}
Again, we perform a finite-size extrapolation to obtain the charge gap in the TL.
Furthermore, we test for the appearance of phase separation (PS) by computing the inverse compressibility, 
\begin{equation}
    \kappa^{-1} = n^2 \frac{\partial^2 e_0}{\partial n^2} \approx n^2\frac{e_0(n+\Delta n) +  e_0(n-\Delta n) - 2e_0(n)}{\Delta n^2} \, ,
    \label{eq:kappa}
\end{equation}
where $e_0=E_0/N$ is the groundstate energy per particle computed for a system with $L$ lattice sites and with $S^z_\text{total}=0$.
We choose $\Delta n = 0.05$, which gives a sufficient approximation to the derivative in Eq.~\eqref{eq:kappa}.

\subsubsection{Correlation functions}
\label{sec:defcorrels}
In order to determine which kind of quasi-long-range order (QLRO) is predominant we compute the correlation functions in the charge, spin, and Cooper channels. 

First, we consider the correlations in total and relative charge density 
\begin{align}\label{eq:ChargeCorrel}
      N(i,j)_{\pm} & = \braket{\hat{n}_i^\pm \hat{n}_j^\pm} - \braket{\hat{n}_i^\pm}\braket{\hat{n}_j^\pm}  ,
\end{align}
 with $\hat{n}_i^\pm = \hat{n}_{i, 1} \pm \hat{n}_{i, 2}$. 
 The corresponding structure factor is computed via
 \begin{equation}
     \mathcal{N}_\pm(k) = \frac{1}{R} \sum_{i,j=1}^{R} \text{e}^{ik(i-j)}N_\pm(|i-j|),
 \end{equation}
 where $R$ is the total number of rungs and $i,j$ denote the $i$th and $j$th rung, respectively.
 
Similarly, we consider correlations of the rung spins 
\begin{equation}\label{eq:SpinCorrel}
        S(i,j) = \braket{S^{z,+}_{i} S^{z,+}_{j}} - \braket{S^{z,+}_{i}} \braket{S^{z,+}_{j}}
\end{equation}
    with $S^{z,+}=S^z_{i, 1} + S^z_{i, 2}$.
    
    Finally, we study {singlet pairing correlations} of rung and leg pair creation/annihilation operators $\Delta_S^{r}(i)=\frac{1}{\sqrt{2}}\left(c_{i, 1, \downarrow}c_{i, 2, \uparrow} - c_{i, 1, \uparrow}c_{i, 2, \downarrow}\right)$, and the leg pairing operator
    $\Delta_S^{l}(i, l)=\frac{1}{\sqrt{2}}\left(c_{i, l, \downarrow}c_{i+1, l, \uparrow} - c_{i, l, \uparrow}c_{i+1, l, \downarrow}\right)$,
    \begin{align}
        P^{rr}_S(i, j) &= \braket{\left(\Delta^{r}_S(i)\right)^\dagger\Delta^{r}_S(j)},\\
        P^{ll}_S(i, j) &= \braket{\left(\Delta^{l}_S(i,l)\right)^\dagger\Delta^{l}_S(j,l)},\\
        P^{rl}_S(i, j) &= \braket{\left(\Delta^{r}_S(i)\right)^\dagger\Delta^{l}_S(j,l)}.
    \end{align}

Since our ladder model is a quasi one-dimensional system, we assume that the correlation functions are predicted by bosonized field theory, i.e., that they are of the form~\cite{Giamarchi2003}
\begin{align}\label{eq:GeneralCorrel}
    C(|i-j|) = \frac{A}{|i-j|^\alpha} + B \frac{\cos(k|i-j| + \varphi)}{|i-j|^\beta} ;
\end{align}
in our numerical approach, the parameters $A\,,B\,,\alpha\,,\beta\,,k\,, \text{and}\,\varphi$ can be used as free parameters to be fitted (they are then supplemented by subscripts $_{c}, _{s}, _{rr},_{ll}, _{rl}$ in the corresponding charge, spin, and Cooper channels).
In the field theory, the values can be expressed using the corresponding LL parameter, see Tab.~\ref{tab:Expectations} and the detailed discussion in the next Sec.~\ref{sec:AnalyticalExpectations}.  
Note that in Eq.~\eqref{eq:GeneralCorrel}, for the sake of simplicity we ignore any kind of logarithmic correction and take only one harmonic into account. 
If one of the pairing correlation functions has the slowest decay, we identify SC quasi-long-range order and call the system superconducting. On the other hand, if either the density or the spin correlations are dominant, we will speak respectively of charge density wave (CDW) and spin density wave (SDW) quasi-long-range order. 

In the SC phases, we distinguish between s-wave and d-wave pairing by means of the relative sign of leg-leg and leg-rung correlators.
However, we emphasize that in the ladder geometry the s- and the d-wave channels correspond to the same irreducible representation of the  symmetries of the system and consequently mix.

\subsubsection{Analytical expectation for correlation functions}
\label{sec:AnalyticalExpectations}

The exponents entering these correlation functions are related to the Luttinger parameter in charge and spin sector, respectively. In table~\ref{tab:Expectations} we summarize the analytical expectations for this relationship. In the following we comment on the results.

\begin{table}[]
    \centering
    \begin{tabular}{c||c|c|c}
         & charge & spin & Cooper \\
         \hline \hline
        $A_{\rm low}$ & $-K_{c,0}/\pi^2$ & $-1/\pi^2$ & $0$\\
        \hline
        $\alpha_{\rm low}$ & $2$ & $2$ & N/A\\
        \hline 
        $\beta_{\rm low}$ & $K_{c,0} + 1$ ($K_{c,0}$) & $K_{c,0} + 1$ (gap) & $1/K_{c,0} + 1$ ($1/K_{c,0}$) \\
        \hline \hline
          $A_{\rm high}$ & $-2K_{c,+}/\pi^2$ & $0$ & $0$\\
        \hline
        $\alpha_{\rm high}$ & $2$ & N/A & N/A\\
        \hline 
        $\beta_{\rm high}$ & gap & gap & $1/2K_{c,+} $\\
        \hline
    \end{tabular}
    \caption{Summary of expected coefficients in the power-law Eq.~\eqref{eq:GeneralCorrel} in the low and high density limit, respectively. All power-laws in the Cooper channels (\textit{rr,rl, ll}) are expected to be equal. The low-density limit displays both a C1S1 phase and a C1S0 phase (results for the latter being quoted in brackets).}
    \label{tab:Expectations}
\end{table}

First, we consider correlations in the charge channel, Eq.~\eqref{eq:ChargeCorrel}. From the definition of fermionic creation and annihilation operators in the band of $0/\pi$ orbitals, Eq.~\eqref{eq:0PiExpansion}, it is evident that $\hat{n}_i^- = c^\dagger_{i, 0} c^{\phantom{\dagger}}_{i,\pi} + \text{H.c.}$ is gapped throughout the phase diagram. In contrast the correlations of the total charge are gapless and we use the field theory expression for the total density \cite{Giamarchi2003}
\begin{align} \label{eq:TotalDensity}
    \hat n_i^+ &\simeq 2n - \sum_{a = 0,\pi} \Big [\frac{\sqrt{2}}{\pi}  \Phi_{a,c}' \notag \\
    &+ 2n [e^{i (2k_F^a x_i - \sqrt{2} \Phi_{a,c})} \cos(\sqrt{2} \Phi_{a,s}) + \text{H.c.}] \notag\\
    &+ 2n [e^{i( 4k_F^a x_i - \sqrt{8} \Phi_{a,c})}  + \text{H.c.}] \Big ].
\end{align}
Because of spin-interactions, the operator for $4k_F$ oscillations is generally expected to be effectively independent of $\Phi_{a,s}$~\cite{Giamarchi2003}. The bosonized expression for the total density has multiple implications for the density correlator Eq.~\eqref{eq:ChargeCorrel}, where we use the form given by Eq.~\eqref{eq:GeneralCorrel}. To begin with, we concentrate on the low density limit $\mu_\pi<-E_p$ below the lower dashed line of Fig.~\ref{fig:PhaseDiagramtJ}. The first term in Eq.~\eqref{eq:TotalDensity} leads to $\alpha_c = 2$ and $A_c = -K_{c,0}/\pi^2$. As long as the spin sector is gapless, both $k = 2k_F^0$ and $k = 4k_F^0$ oscillations are possible and decay with $\beta_c = K_{c,0} + 1$ and $\beta_c = K_{c,0}$, respectively. In contrast, when the spin sector is gapped and $\Phi_{\sigma,0} \in 2\pi \mathbb Z/\sqrt{8}$ (a singlet superconductor), the cosine in the density operator orders and $2k_F^0$ oscillations dominate the result.

The situation is different when $\mu_\pi > E_p$ -- here only $\Phi_{c,+}$ is gapless. While $\alpha_c = 2$ persists, $A_c = -2K_{c,+}/\pi^2$. The density-density correlators do not display power-law correlated oscillations, yet, the correlator of $(\hat n_i^+)^2$ does\cite{WhiteAffleck2002}.

Next, we review the spin channel \cite{Giamarchi2003}. In full analogy to the above said, the $z$ component of the total magnetization is
\begin{align}\label{eq:totalSpin}
    S_{i}^{z,+} &\simeq - \sum_{a = 0,\pi} \Big [\frac{\sqrt{2}}{\pi}  \Phi_{a,s}' \notag \\
    &- 2n [e^{i (2k_F^a x_i - \sqrt{2} \Phi_{a,c})} \sin(\sqrt{2} \Phi_{a,s}) + \text{H.c.}].
\end{align}
In the single band limit $\mu_\pi<-E_p$, we expect two phases. In the C1S1 phase, $\alpha_s = 2$ with $A_s = -1/\pi^2$ and predominant $k = 2k_F^0$ oscillations with $\beta_s = K_{c, 0} + 1$. When the spin sector is gapped and $\Phi_{\sigma,0} \in 2\pi \mathbb Z/\sqrt{8}$, the spin correlator quickly vanishes on length scales larger than the inverse spin gap. This gapped behavior is expected to qualitatively persist to the large density limit $\mu_\pi>E_p$ where we expect a single bosonic mode in the total charge sector.

Finally, we review the Cooper channel, see also Appendix \ref{app:PairCorrelations}. The rung and leg pair annihilators take the bosonized form
\begin{subequations}
\begin{align} 
\Delta_S^r(i) & \simeq  \frac{-i }{\sqrt{2}} \sum_a (-1)^a e^{- i \sqrt{2} \Theta_{a,c}} \cos(\sqrt{2} \Phi_{a,s}), \\
\Delta_S^l(i) & \simeq \frac{-i }{\sqrt{2}} \sum_a  \cos(k_a) e^{- i \sqrt{2} \Theta_{a,c}} \cos(\sqrt{2} \Phi_{a,s}),
\end{align}
\label{eq:bosonziedCooperPair}
\end{subequations}
where $(-1)^a = 1$ ($(-1)^a = -1$) for $a = 0$, ($a = \pi$). Note that the operator content of these bosonized expressions is the same, but the relative sign changes below and above the Lifshitz transition: For low densities, $\cos(k_0) >0$, while for large densities when both bands are populated $\cos(k_a) \propto - (-1)^a$. This accounts for the transition from s-wave to d-wave pairing.

In summary, we expect correlation functions with $A = 0$ for all Cooper correlators (\textit{rr, ll, rl}). We only considered $k = 0$, for which $\beta = 1/K_{c,0} + 1$ in the C1S1 phase and $\beta = 1/K_{c,0}$ in the C1S0 phase. At large densities far above the Lifshitz transition, there is only a C1S0 phase with gapless $\Phi_{c,+}$ mode. Again, $A = 0 = k$ but $\beta = 1/(2K_{c,+})$. In contrast to the low density limit, where $K_{c,0}>1$ implies the dominance of superconducting correlations, for large enough filling the condition becomes $K_{c,+} > 1/2$.

\subsubsection{Determining the Luttinger parameter from the charge structure factor}
\label{sec:LuttParam}

In field theory, Luttinger parameters are defined by means of the prefactor of the action in Eqs.~\eqref{eq:Skin} and \eqref{eq:C1S0}. While the Luttinger parameter of the spin modes is $K_s = 1$ whenever a gapless spin mode is present in our phase diagram, we here briefly summarize how we determine the non-universal Luttinger parameter in the charge sector.

As we find that throughout the phase diagram there is only one gapless charge mode, we numerically obtain the Luttinger parameter in charge space by studying the structure factor of the density correlation functions, Eq.~\eqref{eq:ChargeCorrel}. 
Using the form Eq.~\eqref{eq:GeneralCorrel} and $\alpha_c = 2$, for small values of the momenta $k$ this leads to 
\begin{equation}
    \mathcal N(k) = - \pi A_c \vert k \vert.
    \label{eq:structFactorLimit}
\end{equation}
In the one band (two band) regime $A_c = - K_{c,0}/\pi^2$ ($A_c = - 2K_{c,+}/\pi^2$). 
Hence, the Luttinger parameter can be obtained by fitting Eq.~\eqref{eq:structFactorLimit} for $k \to 0$ to the numerically obtained structure factor. 
The slope is then equal to $-\pi A_c$, and applying the corresponding relation gives the Luttinger parameter.
Throughout the paper, we use the relation for $A_{\rm low}$ to obtain the value of $K_c$.
This is the value displayed in the phase diagrams of Fig.~\ref{fig:PhaseDiagramtJ}.
However, as discussed in Sec.~\ref{sec:fieldtheory}, in the crossover region around the Lifshitz transition we expect the numerical value of the slope to be given by an unknown mixture of $K_{c,0}$ and $K_{c,+}$, so that its interpretation is more involved than in the respective high- or low-density case, and the so-obtained numerical value cannot directly be used to characterize the phase diagram. 
We therefore suppress the index $+$ or $0$ in $K_c$, since this aspect becomes only relevant outside the crossover region.

\subsubsection{Friedel oscillations and precursor Wigner crystal}

In Ref.~\onlinecite{Eggert2009} Friedel-like oscillations were used to identify a precursor region to a Wigner crystal in low-filled Hubbard chains.
The $4k_{F}^0= 4\pi n$ term, which has a similar origin as the $4k_{F}^0 $ in Eq.~\eqref{eq:TotalDensity}, is interaction induced and is prominent at smallest density and values of $J/t$ in Fig.~\ref{fig:PhaseDiagramtJ}.
Here, we apply the same analysis to our $t$-$J$-ladder systems and use the bosonization prediction for the density in the one-band regime~\cite{Eggert2009}
\begin{align}
    n(x_i) &= n - F_1 \frac{\sin(2\pi n x_i)}{[\sin(\frac{\pi x_i}{L + 1})]^{\frac{K_{c,0} + 1}{2}}} - F_2 \frac{\sin(4\pi n x_i - \phi)}{[\sin(\frac{\pi x_i}{L + 1})]^{2K_{c,0}}}.
    \label{eq:FriedelEggert}
\end{align}
It is also possible to estimate the value of the Luttinger parameter by fitting Eq.~\eqref{eq:FriedelEggert} at very low densities.
For the C1S1 phase, the results are consistent with the ones obtained from the structure factor.
However, in the C1S0 phase the fits are much more difficult to control, so that we will not further discuss this approach.

Fitting Eq.~\eqref{eq:FriedelEggert} allows us to identify the precursor region of the Wigner crystal, for which the ratio $F_2/F_1$ is finite, but vanishes outside this region. 
An exemplary fit of the local electron density in the Wigner crystal regime as well as $F_2/F_1$ as a function of $J/t$ are presented in Appendix~\ref{app:further_results}.
As can be seen there, we do not obtain a sharp transition, but a crossover region in which the value $F_2/F_1$ gradually decreases.

\subsection{Details on the MPS calculations}
\label{subsec:DMRG}
We used the MPS-code contained in the SymMPS-package \cite{SymMPS} to calculate the ground state energies, local observables, and correlation functions. 
Both variants of the model were calculated with open boundary conditions on systems with up to 200 lattice sites. 
We used 22 sweeps to ramp up to the maximal bond dimensions $\chi_{\rm max} = 2000$, and afterwards continue with further sweeps until convergence is reached using this value of $\chi_{\rm max}$ (a sweep is going once through the lattice).
This is done by setting the SymMPS control parameter for the ground state energy (which roughly corresponds to the absolute error in the ground state energy \cite{oral_communication_SP}) to $10^{-10}$.
Usually, this threshold is reached after $30-50$ sweeps, in more difficult cases we went up to $200$ sweeps. 
Also in these more difficult cases, a SymMPS-error of the energy $\sim 10^{-9}$ is obtained. 
In addition, the parameters are set such that a discarded weight of $10^{-12}$ is obtained. 
However, in particular at the higher values of the filling treated by us ($n \sim 0.4-0.5$), the entanglement in the system is so high that the discarded weight in the final sweep is $\sim 10^{-6}$. As a reference, it would be interesting to compare our numerical results to approaches using the full SU(2) symmetry of the systems, or by reducing the entanglement, e.g., using the mode transform discussed in Refs.~\onlinecite{legeza1,legeza2}. In the context of our paper, the results shown at the higher values of $n$ will have a larger numerical error, and we will show only the results, for which we have the highest confidence. 

\section{Numerical Results}
\label{sec:Results}

\subsection{Global Observables}

\begin{figure}
	\centering 
	\includegraphics[width=0.45\textwidth]{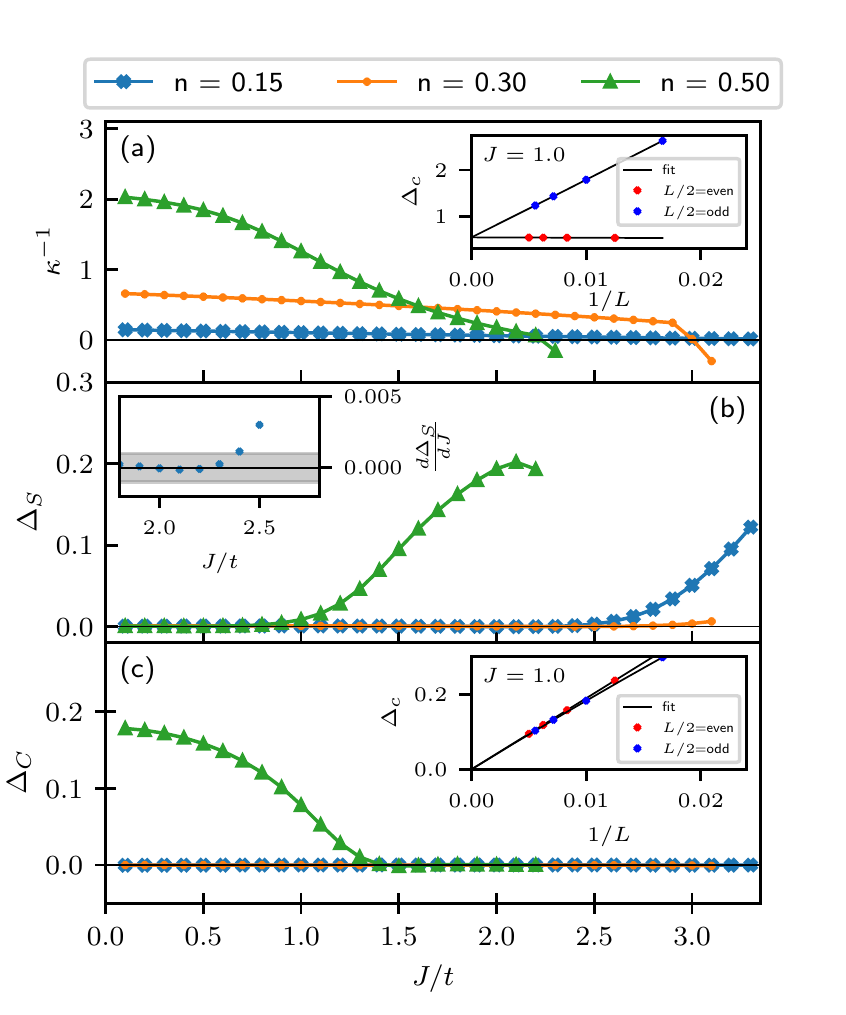}
	\caption{Comparison of the inverse compressibility (top), the spin gap (middle) and the two particle charge gap (bottom) for the regular \tJ ladder in the TL as function of $J/t$ for the densities $n=0.15$ (blue), $n=0.3$ (orange) and $n=0.5$ (green). The inset in the central panel displays the spin gap at $n=0.15$ close to its opening point. It also shows the estimated numerical accuracy of $10^{-3}$, indicated by the gray region. The spin gap is considered to be finite if its value exceeds $10^{-3}$. The inset in the lower and upper panel show, respectively, an example for the finite size scaling of the charge gap and inverse compresibility. Red and blue dots denote the values with $J/t=1$ for systems where the number of lattice sites divided by two is even or odd, respectively.}
	\label{fig:dSinverseKdC_V=1}
\end{figure}

Fig.~\ref{fig:dSinverseKdC_V=1} shows examples for our results for the inverse compressibility as well as for the spin and charge gap of the regular \tJ ladder after extrapolating to the TL as a function of $J/t$ at densities $n=0.15$, $n=0.30$, and $n=0.50$.
The value of $J/t$ at which the inverse compressibility $\kappa^{-1}$ vanishes marks the onset of phase separation. 
For $n=0.50$ the system undergoes the transition to phase separation at $J/t \approx 2.3$ while it happens at $J/t \approx 3.0$ and $J/t \approx 3.4$, respectively, for $n=0.30$ and $n=0.15$. 
Note that in the phase separation region the numerics become unstable due to the high degeneracy of the ground state, so that we do not display results inside the PS region.

The spin gap opens exponentially slowly, which makes it difficult to estimate the exact position of the critical point.  
We make a conservative estimate for the numerical accuracy of the values of the spin gap in the TL to be $\sim 10^{-3}$ (see, e.g.,  \onlinecite{MichaudPRB2010} for an estimate of the accuracy that can be obtained when determining a gap in spin ladders using MPS). 
In Refs.~\onlinecite{Gorshkov1,Manmana2017}, a similar threshold was used to identify the line at which the spin gap opens.
In this way, we estimate at $n=0.50$ the spin gap to open at $J/t \approx 0.7$ while for $n=0.15$ it opens at $J/t \approx 2.4$. 
In the case of the intermediate densities $n=0.30$, the spin gap opens at a later point $J/t \approx 2.8$.
The inset in Fig.~\ref{fig:dSinverseKdC_V=1}b illustrates our approach: it shows the spin gap for $n=0.15$ close to the critical point determined by the spin gap. The estimated error is marked by a gray-colored interval around 0. 
We find the value of the spin gap exceeds the error estimate for $J/t \geq 2.4$ which is, thus, our estimate for the critical point.

The charge gap is zero for almost all investigated densities and all values of $J/t$, within an accuracy of $\sim 10^{-3}$. Only the density $n=1/2$ forms an exception for which a commensurate filling allows for a charge gapped, CDW ordered insulating phase, which we denote by CDWI in order to distinguish from the CDW LL phase. In the regular \tJ model, this phase extends at $n=0.5$ from $J/t=0$ to $J/t \approx 1.5$ while for the $V=0$ case the phase ends at $J/t\approx2.4$.

In order to extrapolate our data into the TL, we performed a finite-size scaling by fitting a second-order polynomial to the numerical results for different system sizes $L$ as a function of $1/L$. Examples of such fits are shown in the insets of Fig.~\ref{fig:dSinverseKdC_V=1}a an Fig.~\ref{fig:dSinverseKdC_V=1}c. Thereby, we encountered an even-odd effect. Every time $L/2$ is an odd number, we artificially remove one electron to ensure that there are as many spin up electrons as there are spin down electrons, and our system stays in the subspace of zero magnetization. Interestingly, the systems sizes for which this was necessary show a different scaling behavior than the other systems. However, in both cases, the numerical results scale to the same value in the TL within an accuracy of $10^{-3}$, consistent with the discussion above.

We compute the central charge according to the procedure outlined in Sec.~\ref{sec:AnalyticalExpectations} and Appendix~\ref{app:entropy} for finite systems up to $L=200$ sites with a subsequent finite-size extrapolation.
We obtain a value for the central charge of $c \approx 2$ in the C1S1 phases and $c \approx 1$ in the C1S0 regions of the phase diagrams of both systems treated. 
This transition is in good agreement with the position of the line at which the spin gap opens.
These findings are in agreement with the results for the spin and charge gaps described before. 
A more detailed discussion of these findings and of the accuracy estimated by us is found in Appendix~\ref{app:entropy}.

\subsection{Luttinger Parameter}

\begin{figure}
    \centering
    \includegraphics{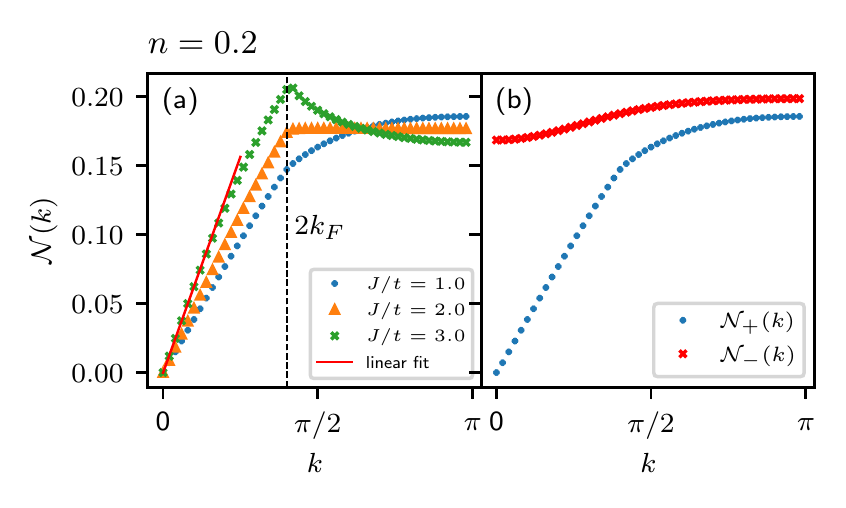}
    \caption{Examples for the structure factor of the total density correlation functions $N_+$ at $n=0.2$ for the regular \tJ ladder for systems with 200 lattice sites and different values of $J/t$. (a) structure factor of $N_+$ for $J/t=1$ (SDW phase), $J/t=2$ (close to the phase transition), and $J/t=3$ (deep in the superconducting phase). The black dashed line denotes the wave-number $2k_F$ for $n=0.2$, while the red solid line is an example of a linear fit to the values corresponding to the six smallest $k$-values. (b) comparison between the structure factor for the $N_+$ and $N_-$ correlation functions for $J/t=1$.}
    \label{fig:exampleStructureFactor}
\end{figure}

Fig.~\ref{fig:exampleStructureFactor}a shows an example for the structure factor of $N(|i-j|)_+$ for different values of $J/t$ at density $n=0.20$. As it can be seen, the structure factor goes linearly to $0$ (with an accuracy  $ \sim 10^{-11}$) for $k\to0$. This is in agreement with the expectations from bosonization and can be used to estimate the value of the Luttinger parameter according to Eq.~\eqref{eq:structFactorLimit}. Furthermore, one can observe that the structure factor develops a kink at $2k_F$ when the system approaches the spin gapped phase. An exception to this behavior forms the structure factor for very small densities and low $J/t$ where we observe a discontinuity in the derivative of the structure factor at $4k_F$, indicating the existence of such oscillations in the correlation function and therefore in the particle density. As Ref.  \cite{Eggert2009} argued, strong interactions between the electrons cause $4k_F$ oscillations, which indicate the appearance of a precursor phase towards a Wigner crystal.  

In Fig.~\ref{fig:exampleStructureFactor}b we compare the structure factors of the $N_\pm(|i-j|)$ correlation functions. The structure factor $\mathcal{N}_-(|i-j|)$ does not vanish for $k\to0$ and has a vanishing derivative at $k=0$. Therefore, in contrast to $N_+(|i-j|)$, the $N_-(|i-j|)$ correlations contain contributions of gapped modes and support the results obtained by the central charge that at least one charge mode must be gapped.  

The Luttinger parameter of the ungapped mode can be obtained by a linear fit of the structure factor for values close to $k=0$ and multiplying the slope with $\pi$ (compare Eq.~\eqref{eq:structFactorLimit}). For our analysis, we waive a finite-size scaling and only use the structure factor of the largest system size we investigated (i.e. $L=200$) since the lattice size seems not to play an important role, as further discussed in Appendix~\ref{app:further_results}.
Even though the structure factors in Fig.~\ref{fig:exampleStructureFactor} seem completely linear close to $k=0$, they still contain slight oscillations. This might be an artifact caused by finite-size effects. Therefore, to average out these oscillations, the fit must be performed through more than just the closest two values to $k=0$ as one would expect. It turns out that using the six closest values to $k=0$ reproduce the results in Ref. \cite{FeiguinAffleck2008} for $n=0.25$ very well, so that we used this for all other cases as well. An exemplary fit is shown in Fig.~\ref{fig:exampleStructureFactor}a.

Usually a Luttinger parameter larger than one indicates attractive interactions in the low-energy field theory and one expects SC. 
However, in the crossover region around the Lifshitz transition indicated by the dashed lines in Fig.~\ref{fig:PhaseDiagramtJ}, this is not reproduced. We associate this discrepancy to the complications discussed in Sec.~\ref{sec:fieldtheory}, so that the characterization of the phase diagram is only possible by carefully analyzing the behavior of correlation functions and the aforementioned global observables.

\subsection{Correlation Functions and Superconductivity}

An important aspect to characterize the different phases is to determine the dominant correlation function, i.e., the algebraically decaying correlation function with smallest exponent. 
Based on the bosonization results discussed in Sec.~\ref{sec:defcorrels}, we use the fit function 
\begin{multline}
    D(|i-j|) = \frac{A}{|i-j|^{\alpha}} +  B\frac{\cos(k_1|i-j| - \phi)}{|i-j|^\beta}\\ + C\frac{\cos(k_2|i-j| - \psi)}{|i-j|^\gamma},
    \label{eq:fit_correlation_functions}
\end{multline}
where we neglected logarithmic corrections, and fit the numerical results for the correlation functions for the total density ($c$), the total spin ($s$), for rung-rung pairing ($rr$), leg-leg pairing ($ll$), and rung-leg pairing ($rl$).
Note that in Eq.~\eqref{eq:GeneralCorrel} we only took one harmonic into account, while in Eq.~\eqref{eq:fit_correlation_functions} we allow for two oscillating contributions with different $k$-values. 
However, it turned out that for all pairing and for the spin correlations, the main contributions in the oscillatory part come from $2k_F$ oscillations. 
If we keep all three terms, the fit routine becomes more unstable even though it mostly finds only $2k_F$ contributions. 
Therefore, these correlations are only fitted with one of the two oscillatory terms, as in Eq.~\eqref{eq:GeneralCorrel}. 
For the fit of the total density correlation function, we used, however, all three terms if also $4k_F$ contributions are relevant, i.e., if we were in the precursor Wigner crystal region.  

We introduce the notation for the smallest (i.e. dominant) exponent of all correlation functions via
\begin{equation}
    x_{\xi} = \text{min}(\alpha_\xi, \beta_\xi \, , \gamma_\xi),\quad \xi\in\{c,s,rr,ll,rl\} \, .
\end{equation}
If the total density or the total spin correlation functions dominate, we consider the system to be in a CDW or SDW phase, respectively. If, on the other hand, one of the pairing correlations dominates, we call the system superconducting.

Note that for the leg-leg pairing correlation functions we only calculated intra-leg correlations. Furthermore, in order to average out possible contributions from the boundary, we apply a running average by choosing a window of 6 sites around the center of the system.
The average is then taken for each possible distance $|i-j|$ over the six different initial positions $i \in [L/2-3,L/2+2]$.  

\begin{figure}
	\centering
	\includegraphics{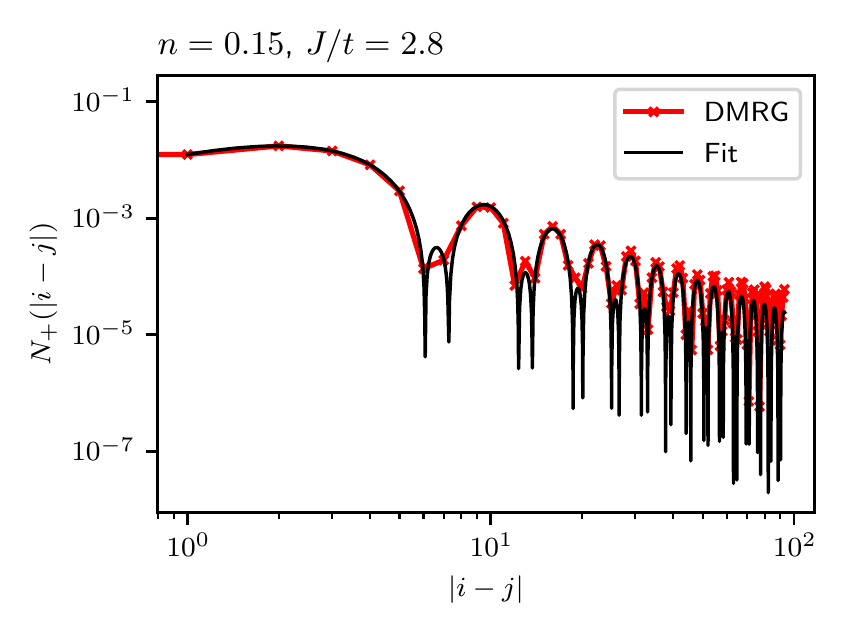}
	\caption{Exemplary fit of the total density correlation function $N_+$ at $n=0.15$ and $J/t=2.8$}
	\label{fig:fit_correlation_function}
\end{figure}

An exemplary fit of the total density correlation function is shown in Fig.~\ref{fig:fit_correlation_function}. 
As can be seen,
it deviates from a purely algebraic decay at large distances.
This can hint to logarithmic corrections, boundary or finite size effects, or difficulties in the convergence of the MPS and shows that it is difficult to obtain the value of the exponents with a high accuracy. Typically, in order to minimize the effect of these artifacts, we restrict our fit to consider only the first 10-30 data points. As further discussed below, we estimate the relative error in our values of the exponents to be between 10-20\%.

\begin{figure}
	\centering
	\includegraphics{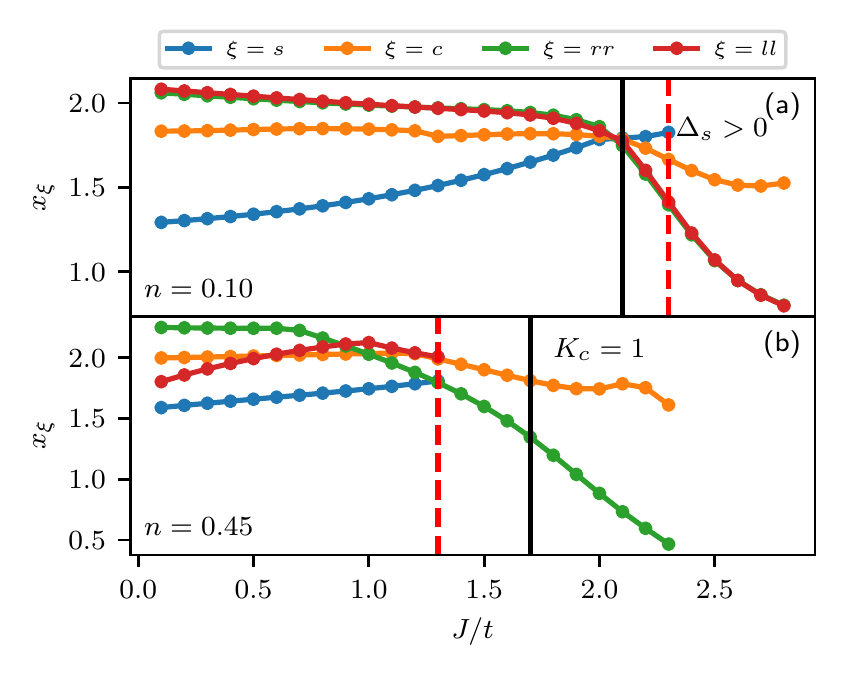}
	\caption{Smallest exponent $x_\xi$ of the total density (orange), the total spin (blue), the rung-rung pairing (green), and the leg-leg pairing (red) correlation functions as a function of $J/t$ for the regular \tJ model. (a) results for density $n=0.10$; (b) results for $n=0.45$. The graphs stop where the system enters phase separation. In addition, the total spin correlation function stops where the system develops a spin gap. The black line denotes the value of $J/t$ for which the Luttinger parameter $K_c = 1$, determined by the structure factor of the total density correlation function. The dashed red line indicates where the spin gap opens and the spin correlations start to decay exponentially. 
	}
	\label{fig:smallest_expo}
\end{figure}

Fig.~\ref{fig:smallest_expo} shows our results for the regular \tJ model for $x_\xi$ as a function of $J/t$ for different correlation functions at densities $n=0.1$ and $n=0.45$. 
We do not show the exponents for the rung-leg correlation functions since the fits are numerically unstable and give different qualitative behavior depending on the number of oscillatory terms we keep. 
This is further discussed in Appendix~\ref{app:further_results}, where we find the Fourier spectrum of these correlation functions not to contain sharp $2 k_F$ oscillations, in contrast to the other pairing correlation functions, so that is much more difficult to obtain meaningful fits using expression~\eqref{eq:fit_correlation_functions}. 
Note that for $n=0.45$ we show our results for $x_{ll}$ only in the C1S1 phase. For larger values of $J/t$, we find the fits to become unstable. This can be due to the lower accuracy obtained in this region of the phase diagram, as discussed in Sec.~\ref{subsec:DMRG}.

Fig.~\ref{fig:smallest_expo}a
demonstrates that below the crossover region of the Lifshitz transition the expectations from bosonization are fulfilled: for $K_c<1$, the spin correlations decay slowest, while for $K_c>1$ the pairing correlations dominate, irrespective of the presence of a spin-gap.
This is the same behavior as observed for chains at low fillings \cite{Moreno2011,Manmana2017}. 
In addition, the different pairing channels are degenerate.

In contrast, at $n=0.45$ the
value of $J/t \approx 1.7$ at which $K_c=1$ and the value of $J/t \approx 1.3$, at which the pairing correlations become dominant, clearly disagree.
The pairing correlations become dominant as soon as the spin gap opens.
This further illustrates that in the vicinity of the Lifshitz transition the value of $K_c$ as obtained by us cannot be used to directly characterize the phase diagram.
For small $J/t$ the non-oscillating term in the density correlation function dominates ($\alpha_c$) at both values of the filling shown. 
According to Tab.~\ref{tab:Expectations}, one expects $x_c = 2$. 
While for $n=0.45$ this is obtained in good accuracy, at $n=0.1$ we find $x_c\sim 1.85$, clearly deviating from the expected value.
This allows us to estimate a relative error of $\sim 10\%$ for this quantity, and in a conservative estimate we assume the relative error for all exponents shown in Fig.~\ref{fig:smallest_expo} to be of the order $\lesssim 20\%$ 
The phase boundaries shown in Fig.~\ref{fig:PhaseDiagramtJ} are determined using this approach, and hence are affected by a small error bar, which is not displayed there for the sake of clarity.

\begin{figure}
	\centering
	\includegraphics[width=0.45\textwidth]{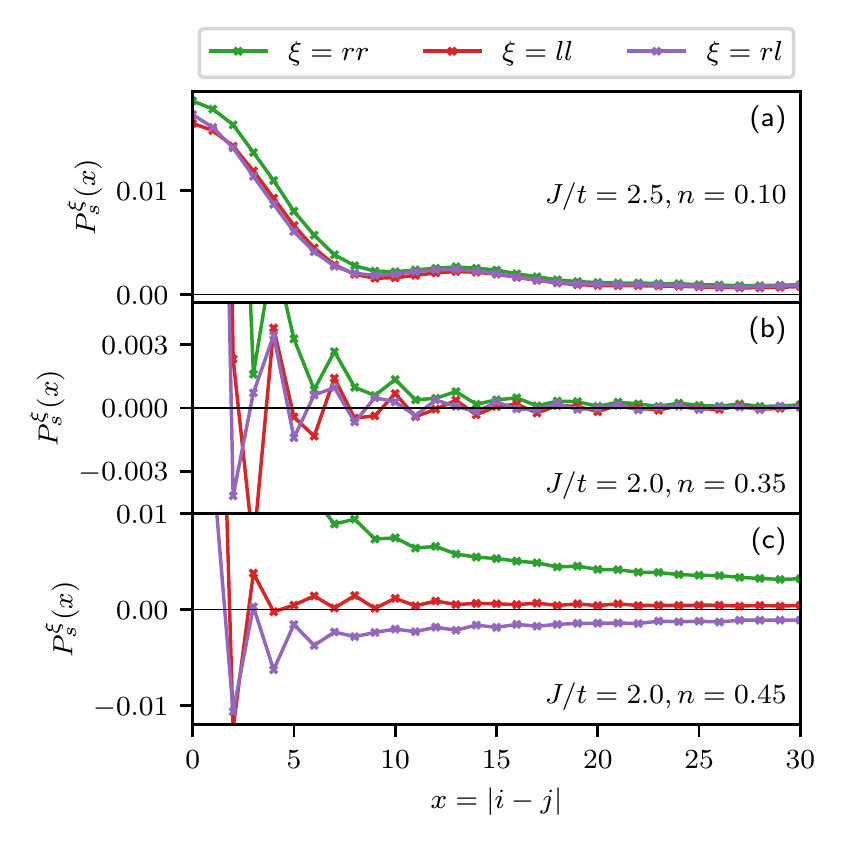}
	\caption{Comparison of the different pairing correlation functions at parameters $n=0.10$ and $J/t=2.5$ (a), $n=0.35$ and $J/t=2.0$ (b), and $n=0.45$ and $J/t=2.0$ (c) in the superconducting phase of the regular \tJ model.}
	\label{fig:dwave}
\end{figure}

We distinguish between s-wave-like and d-wave-like pairing based on the discussion in  Ref.~\onlinecite{Noack1994}. We find that in the superconducting phases, the rung-rung correlation functions are strictly positive. Thus, the sign of the rung-leg pairing  reveals if a rung and a leg pair have a relative phase. If $P_s^{rl}$ has the same sign as $P_s^{rr}$ (i.e. positive), we call the phase (extended) s-wave SC, while for different sign (i.e. negative values of $P_s^{rl}$) we call the phase d-wave SC. This classification is inspired by the definitions of s- and d-wave SC on a square lattice.
As can be seen in Fig.~\ref{fig:dwave}, the relative sign between $P_s^{rr}$ and $P_s^{rl}$ depends on the filling $n$. 
At low filling ($n=0.1$), both have positive sign.
At intermediate filling ($n=0.35$), $P_s^{rr}$ remains strictly positive, but $P_s^{rl}$ oscillates around zero.
At higher filling ($n=0.45$), we see that for large distances $|i-j|$ $P_S^{rr}$ and $P_S^{rl}$ have a different sign for all distances, 
illustrating the crossover from s-wave like to d-wave like SC when increasing the densities. 
	
\subsection{Gas of electron pairs}

\begin{figure}
    \centering
    \includegraphics{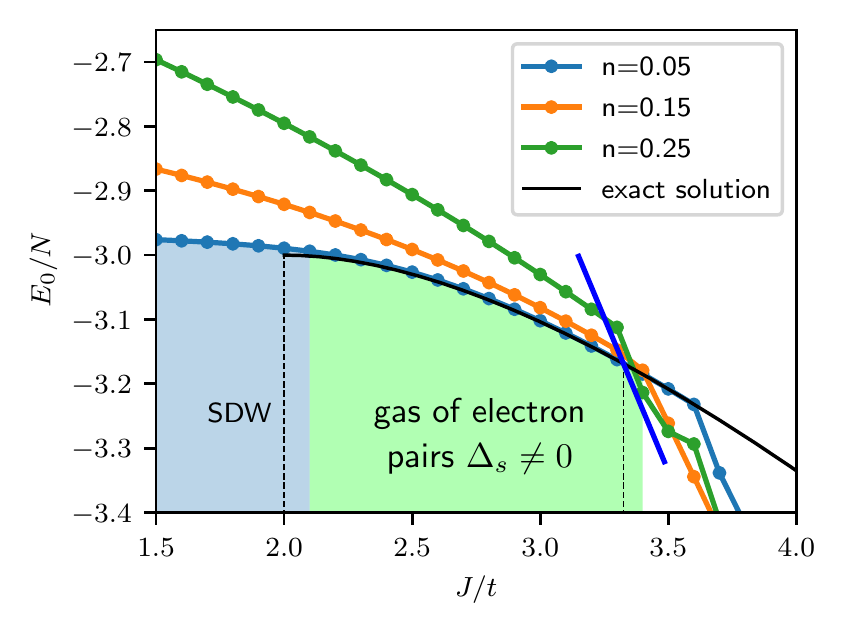}
    \caption{Groundstate energy per particle as a function of $J/t$ for the densities $n=0.05$, $n=0.15$, and $n=0.25$ in a ladder with 200 sites as a function of $J/t$. The black line is the exact groundstate energy per particle of two particles on an empty infinite lattice. The blue diagonal shows the groundstate energy per particle of the Heisenberg part of the regular \tJ ladder for $n=1$. The left and right dashed lines mark respectively the onset of the spin gapped phase and phase separation predicted by the low density limit. The blue and green array blow the blue graph denotes the SDW and spin gap phase for $n=0.05$ respectively.}
    \label{fig:gas_electron_pairs}
\end{figure}

The formation of a spin gap at very low densities can be explained by electrons forming a \textit{gas of free electron pairs}. We follow the description of such a gas given in Ref. \cite{Moreno2011}. Fig.~\ref{fig:gas_electron_pairs} shows a comparison of the ground state energy per particle for different densities and for the exact solution of a single two-particle bound state (cf. Sec.~\ref{sec:zero_density_limit}). For $n=0.05$ both energies match well in the spin gapped region. This suggests that at $n=0.05$ the groundstate consists of a set of bound electron pairs that are strongly diluted.
Note, however, that the dashed lines show the transitions in the zero-density limit and the deviations already indicate weak interaction corrections.
The non-interacting pair gas picture clearly breaks down at higher densities.
For $n=0.15$, one can argue that the qualitative behavior of the ground state energy as a function of $J/t$ is still consistent with the low energy picture but with an already strong energy shift.

\section{Zero-density limit\label{sec:zero_density_limit}}
In order to obtain a more detailed analytical picture of the lower edge of the phase diagram, we revisit an energetic argument first put forward by Emery et al. \cite{Emery1990}.
It relies on the exact expression for the energy of a two-particle bound state \cite{Lin1991,Ogata1991,Hellberg1995}.
The argument goes like this:
Consider the ground state of two electrons in an infinite $t$-$J$ system.
For small $J/t$ the electrons behave like free particles up to a value of $J_c^{(1)}/t$, above which a bound two-particle state with energy $E_B$ becomes energetically favorable.
In the (very) dilute limit and under the assumption that bound states of a higher number of particles play no role, this yields the boundary from SDW to the ``gas of electron pairs'' at the bottom of the phase diagram in Fig.~\ref{fig:PhaseDiagramtJ}.
The transition point to phase separation, $J_c^{(2)}/t$,  is obtained from comparing the energy of two electrons in a hole-free ``island'' to the singlet bound state.
Since in the TL these islands will still be very large,
one can use the average electron energy of a Heisenberg model ground state.
However, similar to the treatment in the polar molecule quantum simulators~\cite{Gorshkov1,Manmana2017}, we allow in the following also for an $XXZ$-anisotropy,
\[
\hat{H}_{XXZ} = \frac J 2 \sum_j \left(S_j^+ S_{j+1}^- + \text{H.c.}\right) + J_z \sum_j S_j^z S_{j+1}^z \, ,
\]
and introduce the anisotropy parameter $\alpha = J_z/J$.

In one spatial dimension the bound state energy of the conventional $t$-$J$ model is given by $E_B^\text{1d} = -J - 4 t^2/J$.
Adapting the 2d calculation by Lin \cite{Lin1991} to the ladder geometry (cf. Appendix~\ref{app:two_el_solution}) yields an implicit equation of the form
\begin{align}\begin{split}
  -\frac{12t^2}{E_s} = \frac{1}{I_0(E_B)} - E_B,
  \label{eq:two_el_sol_energy}
\end{split}\end{align}
where $E_s$ is the energy of a singlet and
\begin{align}\begin{split}
 I_0(E) \quad &\overset{L \rightarrow \infty}{=} \quad - \frac{1}{2} \Big( \frac{1}{\sqrt{(E-2t)^2 - (4t)^2}} \\
 &\qquad\qquad + \frac{1}{\sqrt{(E+2t)^2 - (4t)^2}} \Big) .
 \label{eq:I_0_ladder}
\end{split}\end{align}
We obtain $J_c^{(1)}$ upon equating $E_B \overset{!}{=} -6t$, i.e. the kinetic energy of two free electrons.
As $I_0(E_B \rightarrow -6t)$ diverges, we need to solve $-12 t^2 / E_s = -E_B = 6 t$.
This results in $E_s \overset{!}{=} -2t$. In the general case with anisotropy, the energy of a singlet is given by $E_s = \big( - \frac{1}{2} - \frac{\alpha}{4} -\frac{V}{4} \big) J$ yielding
\begin{equation}
 J_c^{(1)} = \frac{8 t}{2 + \alpha + V} \, .
\end{equation}
For the conventional $t$-$J$ model this yields $J_c^{(1)} = 2 t$.
We note that the same formula holds true in extended 1d and 2d systems.
The transition point to phase separation, $J_c^{(2)}$, is obtained from the comparison of $E_B$ with the energy of two electrons in the ground state of the hole-free model.
In this case the kinetic term does not contribute and the $V$-term only shifts the energy.
We therefore calculated the ground state energy of the spin-exchange term numerically with MPS.
The dark blue line in Fig.~\ref{fig:gas_electron_pairs} shows the resulting energy of two electrons in the phase separation,
while the black line is the energy of the two-particle bound state \eqref{eq:two_el_sol_energy}.
Hence, $J_c^{(2)}$ is the point where both lines intersect.
For values of $J$ between $J_c^{(1)}$ and $J_c^{(2)}$, it is most favorable for the electrons to form bound pairs giving rise to the gas of electron pairs in the phase diagram.
The numerical results for density $n = 0.05$ are consistent with this prediction but show already some renormalization of the $J_c$.

\begin{figure}
 \includegraphics[width=0.5\textwidth]{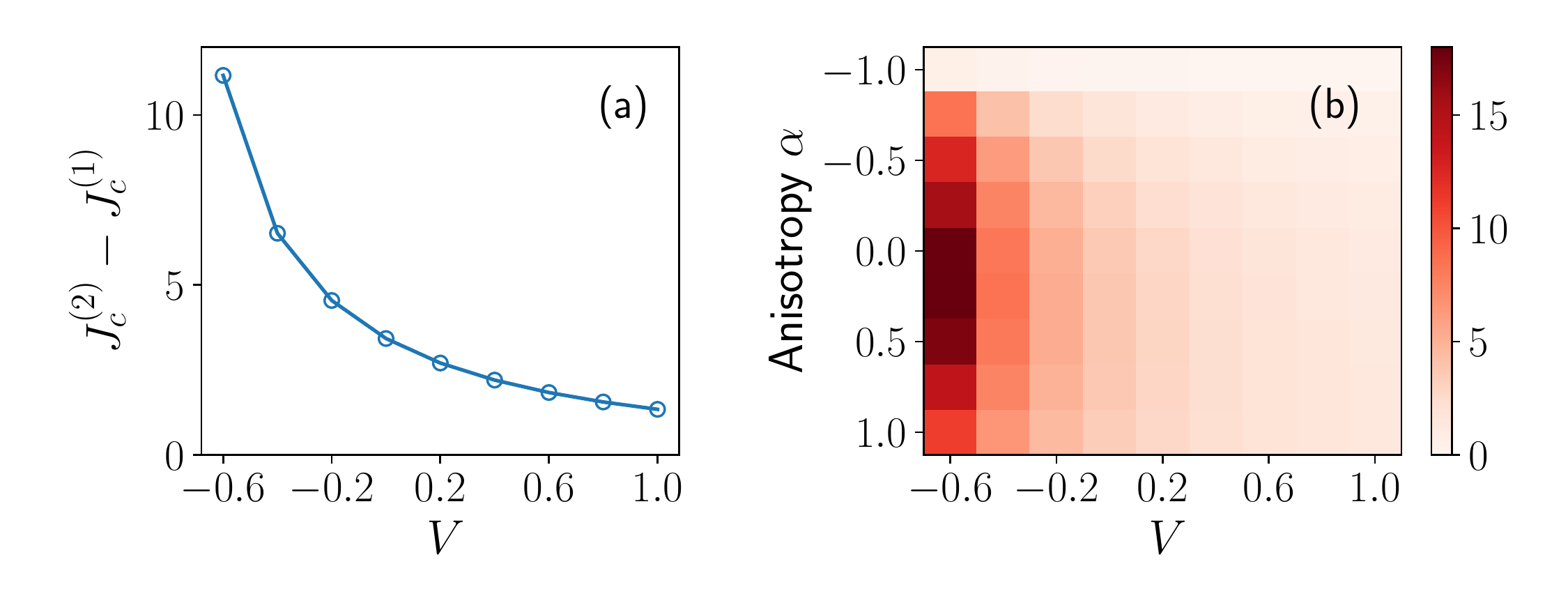}
 \caption{Width $J_c^{(2)} - J_c^{(1)}$ of the superconducting region in the phase diagram in the zero-density limit. a) anisotropy $\alpha = 1$ (standard case in this paper); b) width as a function of both $\alpha$ and $V$. A reduced value of $\alpha$ increases the superconducting region at negative $V$.\label{fig:Jc2minJc1_alpha_beta}}
\end{figure}

In Figure \ref{fig:Jc2minJc1_alpha_beta} we plot the width $J_c^{(2)} - J_c^{(1)}$ of the superconducting region in the phase diagram for varying values of $V$ and the anisotropy $\alpha$.
At $\alpha = 1$ (left plot) the lower edges of the  phase diagrams in Fig.~\ref{fig:PhaseDiagramtJ} are well reproduced.
One can clearly see that a low or negative value of $V$ enhances the superconducting parameter range.
In addition, a value $0 < \alpha < 1$ can also broaden it further.

\section{Summary}
\label{sec:Summary}

Using a combination of numerical MPS and analytical methods, we investigate the ground state phase diagram of two variants of the two-leg \tJ ladder for densities below $n=0.5$. 
The difference between both variants is the strength of the nearest-neighbor Coulomb repulsion $V$, which is known from previous work on chains \cite{Troyer1993,Gorshkov1,Manmana2017} 
to enhance SC phases.

We numerically compute the spin and charge gaps, and the inverse compressibility, which we extrapolate to the TL. 
In addition, we compute for finite systems with up to L=200 sites the central charge, the Luttinger parameter, and correlation functions. 
We determined the Luttinger parameter by calculating the density correlation structure factor.
We complement this by an analytical treatment of the band-filling Lifshitz-type transition using a simple Hartree-Fock ansatz.
At $J/t=0$, the Lifshitz transition, where the Fermi surface changes from two to four Fermi points, happens at $n=0.5$. Using the Hartree-Fock approximation, we estimated the position of the Lifshitz transition as a function of $J/t$.
This band-filling transition is meaningful for the phase diagram of both models, it manifests itself as a crossover between two effective field theories in which the physical meaning of the bosonic charge mode and of the Luttinger parameter is of different nature.
Around this Lifshitz transition, we estimate the size of the crossover region using RPA. While a field theory at the Lifshitz transition~\cite{Meng2011} had been developed before, important predictions such as the crossover function of the Luttinger parameter are still unknown.

For example, at low densities, the numerically obtained Luttinger parameter 
is in good agreement with the behavior of correlation functions and the phase boundary or crossover between phases, as known for \tJ chain systems \cite{Moreno2011,Manmana2017}.
In contrast, while in the crossover region it is still possible to numerically obtain values for a
Luttinger parameter, its behavior and phase boundaries are incompatible with standard bosonization theory. 
We associate this to the aforementioned unknown crossover between 
the two different field theories at large and low densities, respectively, which makes it difficult to obtain meaningful numerical results for the characterization of the phase diagram. 
Therefore, we analyze in detail the decay of different correlation functions in order to characterize the phase diagram.

For small $J/t$, both variants of the \tJ ladder possess dominant SDW correlations in a gapless C1S1 phase. 
Within this phase, at low fillings and for small $J/t$, we identify a precursor of a Wigner crystal, similar to the findings in Hubbard chains~\cite{Eggert2009}. 
For larger values of $J/t$ both systems develop a spin gapped phase (C1S0). 
This observation is supported by the computed value of the central charge, which drops from two to one once the C1S0 phase is entered. For the lowest densities, the spin gapped phase can be understood as a gas of free electron pairs. 
The system with $V=0$, however, has an additional CDW LL phase in the vicinity of the Lifshitz transition before opening the spin gap.
Such a significant change of the phase diagram was not observed in chains \cite{Gorshkov1,Manmana2017}.
This opens the question 
if and how changing $V$ in higher dimensional \tJ-systems (e.g. broader ladder systems or in 2D) can lead to new features in the phase diagram.

The crossover region from s-wave-like SC, in which the sign of rung-rung and rung-leg pairing correlators is the same, to a d-wave-like SC phase in which these correlators have different sign, falls into the crossover regime of the Lifshitz transition. This raises the question if in a more accurate treatment of the Lifshitz transition a connection between both could be seen. This could be investigated, e.g. by computing the electronic spectral functions of the model as a function of the filling, which will allow to determine the precise point of the band filling transition.

One key observation of this paper is that the superconducting phase and the spin gapped phase seem to coincide in the $n\rightarrow0$ limit. 
Setting $V=0$ leads to an enhanced SC phase, whose width can be obtained in this limit in an exact calculation of the binding energy of two electrons and a comparison between the exact ground state energy and the ground state energy of a Heisenberg chain. 
In addition, we exemplarily show for the lower edge of the phase diagram how an XXZ-anisotropy in the spin exchange can modify this phase.
Both, tuning $V$ and the XXZ-anisotropy can be relevant for experiments with ultracold polar molecules on optical lattices, which should allow full tunability of the interactions in the \tJ model. 

It will be interesting to study the Lifshitz transition and its impact on the phase diagram when increasing the number of legs in ladder systems. 
There might be multiple crossovers at which the system may qualitatively change its behavior. Also, it will be interesting to study such broad ladder systems for the Hubbard model or for  magnetically frustrated systems like ladders with next-nearest-neighbour interactions. 

\acknowledgements
We acknowledge insightful discussions with R.M. Noack, \"O. Legeza, D. Chakraborty, A. Schnyder, J. Mitscherling, P. M. Bonetti, D. Vilardi, R. Scholle and V. Mishra.
S.R.M. acknowledges financial support by ESI Vienna, where part of this work was accomplished, and by the Deutsche Forschungsgemeinschaft (DFG, German Research Foundation) - 217133147/SFB 1073, project B03. A.O. acknowledges financial support by the Deutsche Forschungsgemeinschaft (DFG, German Research Foundation) - 217133147/SFB 1073, project B07.

\bibliography{tJMain}

\appendix
\section{Analytical Phase diagram}
\label{app:HartreeFock}

In this appendix we present details on the interaction induced suppression of the Lifshitz transition, as summarized in Sec.~\ref{sec:Lifshitz} of the main text.

In all of the following, we will use the expansion in 0 and $\pi$ orbitals accross the rungs, Eq.~\eqref{eq:0PiExpansion}, to find
\begin{equation}
    \contraction{}{c^\dagger}{_{i,l,}}{c}
    c^\dagger_{i,l,\sigma}c_{i, l', \sigma'}  = (-1)^{l + l'} n \delta_{\sigma \sigma'}/2,
\end{equation}
under the assumption that we are below the Lifshitz transition and thus $\langle \sum_\sigma c^\dagger_{i,0,\sigma}c_{i, 0, \sigma} \rangle = 2 n$ with $n$ the density per site.

We start by studying the Hubbard U term, Eq.~\eqref{eq:HubbardU}. The calculation of the Hartree-Fock self-energy leads to (normal ordering is taken before contractions)
\begin{align}
    \mathcal{H}^{\rm eff}_U & = U \sum_{i,l,\sigma,\sigma'} :\contraction{c^\dagger_{i,l,\sigma}}{c}{_{i, l, \sigma}}{c}
    c^\dagger_{i,l,\sigma}c_{i, l, \sigma} c^\dagger_{i,l,\sigma '}c_{i,l,\sigma '} :\notag \\
    &+U \sum_{i,l,\sigma,\sigma'} : \contraction{}{c^\dagger}{_{i,l,}}{c}
    c^\dagger_{i,l,\sigma}c_{i, l, \sigma} c^\dagger_{i,l,\sigma '}c_{i,l,\sigma '}: \notag \\
    & = \frac{U n}{2}\sum_{i, l, \sigma}  c^\dagger_{i,l,\sigma}c_{i, l, \sigma}.
\end{align}
We employ Einstein summation convention for spin indices.
Note that we neglect possible mean-field decoupling in the spin and pairing channels.
Thus, the Hubbard $U$ term only leads to an overall shift in the chemical potential for both bands. This effect is trivial and thus disregarded in the main text.

Next, we study the term stemming from intrarung interactions 
\begin{align}
    \mathcal{H}^{\rm eff}_{\rm rung} & = \frac{J}{4} [\vec \sigma_{\sigma_1, \sigma_1'} \cdot \vec \sigma_{\sigma_2, \sigma_2'} - V \delta_{\sigma_1, \sigma_1'}  \delta_{\sigma_2, \sigma_2'}] \notag \\
    &\times \sum_i\Big [ \left ( :\contraction{c^\dagger_{i,1,\sigma_1}}{c}{_{i, 1, \sigma_1'}}{c}
    c^\dagger_{i,1,\sigma_1}c_{i, 1, \sigma_1'} c^\dagger_{i,2,\sigma_2}c_{i,2,\sigma_2 '} : + 1 \leftrightarrow 2 \right) \notag\\
    & +\left (:\contraction{}{c^\dagger}{_{i,1,\sigma_1}}{c}
    c^\dagger_{i,1,\sigma_1}c_{i, 1, \sigma_1'} c^\dagger_{i,2,\sigma_2}c_{i,2,\sigma_2 '} : + 1 \leftrightarrow 2 \right) \Big ] \notag \\
    & = -\frac{J V n}{4} \sum_{i,l} c^\dagger_{i, l, \sigma}c_{i, l, \sigma} \notag \\
    &+\frac{J n}{8} \left [ 3 - V\right]\sum_{i} \left (c^\dagger_{i, 1, \sigma}c_{i, 2, \sigma} + \text{H.c.} \right) .
\end{align}
Apart from yet another contribution to the overall chemical potential, this result is the derivation of Eq.~\eqref{eq:tperpRenorm}.

Finally, we evaluate nearest neighbor intrachain interactions,
\begin{align}
    \mathcal{H}^{\rm eff}_{\rm intra} & = 
    \frac{J}{4} [\vec \sigma_{\sigma_1, \sigma_1'} \cdot \vec \sigma_{\sigma_2, \sigma_2'} - {V} \delta_{\sigma_1, \sigma_1'}  \delta_{\sigma_2, \sigma_2'}] \sum_{i,l}\notag \\
    &\Big [ \left ( :\contraction{c^\dagger_{i,l,\sigma_1}}{c}{_{i, l, \sigma_1'}}{c}
    c^\dagger_{i,l,\sigma_1}c_{i, l, \sigma_1'} c^\dagger_{i+1,l,\sigma_2}c_{i+1,l,\sigma_2 '} : + i \leftrightarrow (i + 1) \right) \notag\\
    & +\left (:\contraction{}{c^\dagger}{_{i,l,\sigma_1}}{c}
    c^\dagger_{i,l,\sigma_1}c_{i, l, \sigma_1'} c^\dagger_{i,l,\sigma_2}c_{i,l,\sigma_2 '} : + i \leftrightarrow (i + 1) \right) \Big ] \notag \\
    & = -\frac{J V n}{4} \sum_{i,l} c^\dagger_{i, l, \sigma}c_{i, l, \sigma} \notag \\
    &-\frac{J}{4}\mathcal C \left [ 3 - {V}\right]\sum_{i,l} \left (c^\dagger_{i, l, \sigma}c_{i+1, l, \sigma} + \text{H.c.} \right) ,
\end{align}
where
\begin{align}
    \mathcal C & = \frac{1}{4} \sum_{l, \sigma} \langle c^\dagger_{i,l, \sigma}c_{i+1,l, \sigma} \rangle  = \frac{1}{2} \int_{-k_{F,\pi}}^{k_{F,\pi}} \frac{dk}{2\pi} e^{i k} \notag \\
    &= \frac{\sin(k_{F,\pi})}{2\pi} = \frac{\sin(\pi n)}{2\pi}.
\end{align}
This is the origin of Eq.~\eqref{eq:tPararenorm} of the main text.

\begin{figure}
	\centering
	\includegraphics[]{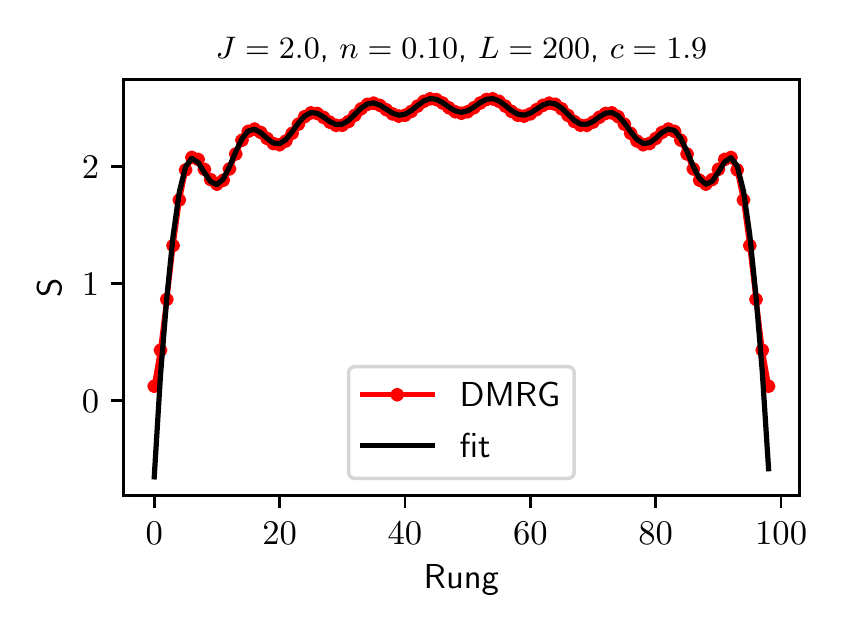}
	\caption{Exemplary fit of the entanglement entropy in the regular \tJ ladder at $n=0.10$ and $J/t=2.0$. The Plot only shows the entanglement entropy for bisections between the rungs.}
	\label{fig:example_fit_c}
\end{figure}

\section{Entanglement entropy and central charge}
\label{app:entropy}
Since we treat systems with OBC, for which the MPS is more efficient, we need to take into account oscillatory terms in the entanglement entropy.
A typical example is shown in Fig.~\ref{fig:example_fit_c}, together with the fitting as explained in Sec.~\ref{sec:fieldtheory}. 
In the shown case, a good quality of the fit is obtained for the largest system size treated by us ($L=200$), and the numerical value of the central charge is $c \approx 1.9$, which is close to the expected value $c=2$.

\begin{figure}[b]
	\centering
	\includegraphics[]{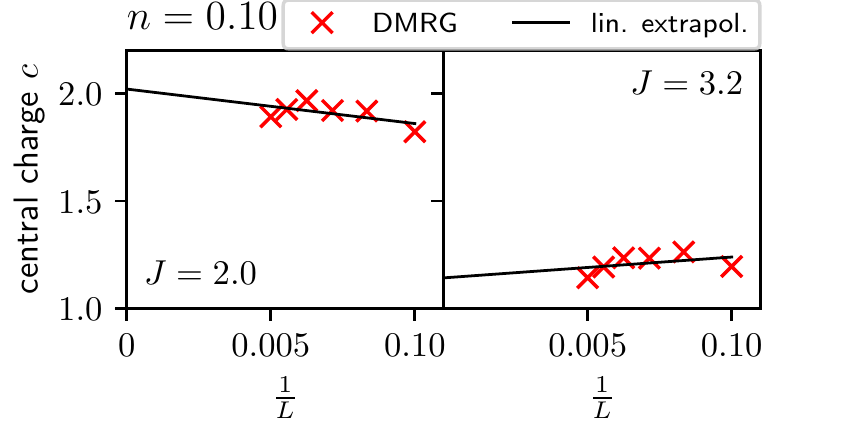}
	\caption{Examplary finite size scaling for the central charge in the regular \tJ ladder. The right panel presents a finite size scaling in the C1S1 phase while the left panel shows a finite size scaling in the C1S0 phase.}
	\label{fig:example_scaling_c}
\end{figure}

In Fig.~\ref{fig:example_scaling_c} we present a finite size extrapolation of the obtained values for $c$ in the C1S1 and the C1S0 phases, respectively.
We see that in both cases finite-size effects are rather small, and that an extrapolation leads closer to the expected values $c=2$ and $c=1$, respectively.
However, in particular for the largest system sizes, a deviation from the linear extrapolation is seen.
This can have different reasons, one of them being the convergence being more difficult to control for the ladder system than for chains, in which this approach was used to obtain the central charge with high precision.
Note also that the oscillations complicate the fitting procedure; it would be preferable to use periodic boundary conditions, in which these oscillatory terms do not appear. 
However, this case is more difficult to control with MPS, in particular also for the ladder systems treated here.
It would be desirable to improve the MPS approach, e.g. using the mode transform of Refs.~\onlinecite{legeza1,legeza2}, in order to reach a higher accuracy in the vicinity of the Lifshitz transition.  
Here, a small error margin remains in our treatment.
Therefore, our results in the TL do not show a sharp drop in the value of $c$ from 2 to 1 at the phase boundary, but more a smooth transition.
Also note the exponentially small opening of the spin-gap, which indicates a large length scale in the vicinity of the critical points.
This also influences the accuracy in the determination of the value of $c$, and much larger system sizes would be needed to obtain the central charge with higher precision, which again is difficult to achieve using MPS for the present ladder systems, so that we refrain from doing so at this point. 

\begin{figure}
	\centering 
	\includegraphics[]{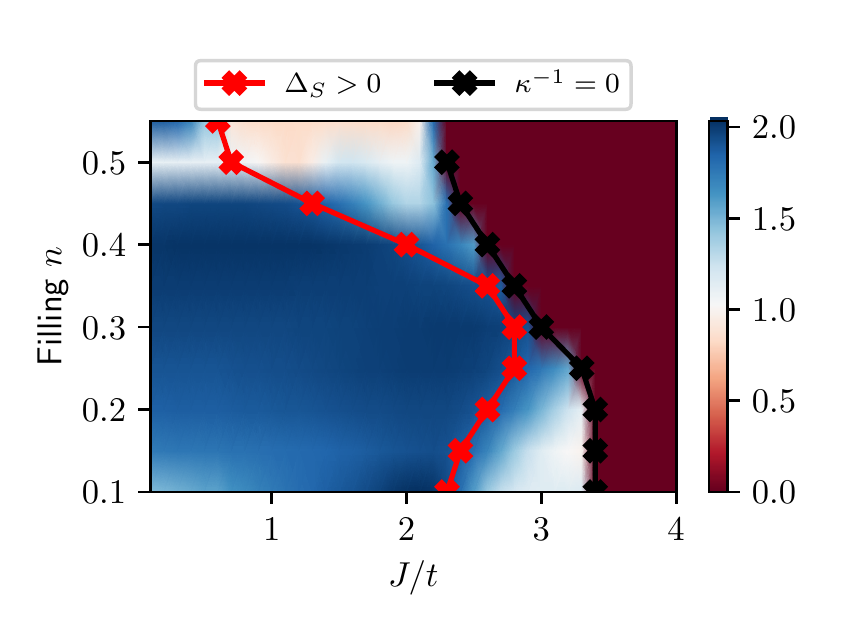}
	\includegraphics[]{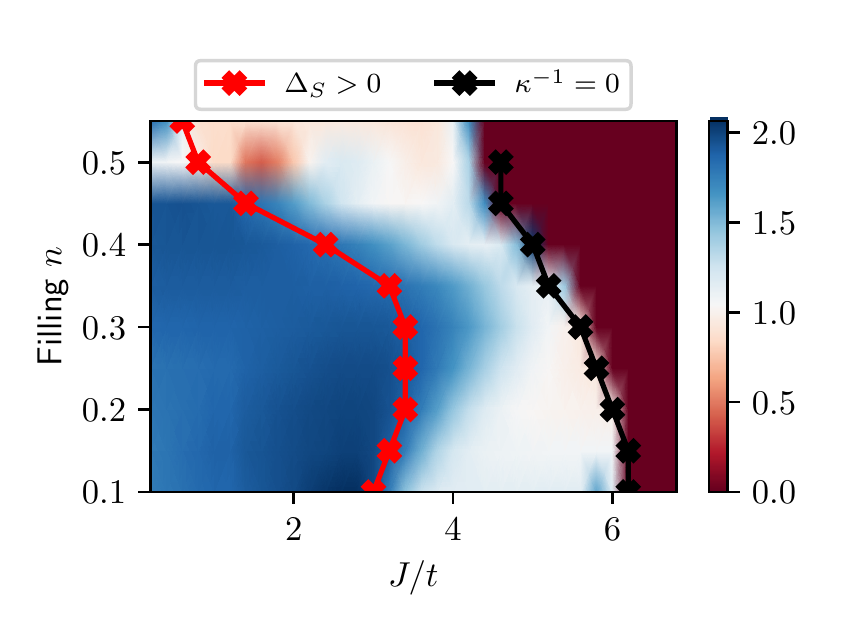}
	\caption{Colormap plot of the central charge in the regular (top) and $V=0$ (bottom) two-leg \tJ ladder. The red line denotes where the spin gap becomes finite while the black line marks the onset of phase separation. In the phase separated region, the numerical data was set to zero. The filling $n=0.05$ is left out since the fits were difficult to control.
	}
	\label{fig:CentralCharge}
\end{figure}

Fig.~\ref{fig:CentralCharge} displays the so-obtained values of the central charge for both phase diagrams shown in Fig.~\ref{fig:PhaseDiagramtJ}. 
Note the difference between the region in which $c=1$ and the line of opening of the spin-gap; while we cannot fully rule out additional effects playing a role, the aforementioned difficulties in reaching a higher precision in the estimation of $c$ make it plausible that throughout the phase diagrams in the TL either a value of $c=2$ (C1S1 phase) or $c=1$ (C1S0 phase) is realized, and that at the phase boundaries this value shows a sharp drop in the TL, in accordance with the expectations from the field theory of Sec.~\ref{sec:fieldtheory}. 

\section{Fourier Transformation of Pairing Correlation Functions}
\label{app:PairCorrelations}

In this appendix we summarize details about the pairing correlation functions quoted in Sec.~\ref{sec:AnalyticalExpectations}.

The rung pair annihilator is 
\begin{align}
    \Delta_S^r(i) 
    & = \frac{-i}{\sqrt{2}^3} \left (c_{i,1}^T,c_{i,2}^T\right)\sigma_y \left(\begin{array}{cc}
     0 & 1 \\
    1 & 0
    \end{array}\right) \left (\begin{array}{c}
         c_{i,1}  \\
         c_{i,2} 
    \end{array} \right) \notag\\
    & = \frac{-i}{\sqrt{2}^3} \left (c_{i,\pi}^T,c_{i,0}^T\right)\sigma_y \left(\begin{array}{cc}
     -1 & 0 \\
    0 & 1
    \end{array}\right) \left (\begin{array}{c}
         c_{i,\pi}  \\
         c_{i,0} 
    \end{array} \right).
\end{align}
In contrast, the symmetric leg-pairing operator is
\begin{align}
   \sum_l\frac{1}{2} \Delta_S^l(i) 
    & = \frac{-i}{\sqrt{2}^3} \left (c_{i,1}^T,c_{i,2}^T\right)\sigma_y \left(\begin{array}{cc}
     1 & 0 \\
    0 & 1
    \end{array}\right) \left (\begin{array}{c}
         c_{i,1}  \\
         c_{i,2} 
    \end{array} \right) \notag\\
    & = \frac{-i}{\sqrt{2}^3} \left (c_{i,\pi}^T,c_{i,0}^T\right)\sigma_y \left(\begin{array}{cc}
     1 & 0 \\
    0 & 1
    \end{array}\right) \left (\begin{array}{c}
         c_{i,\pi}  \\
         c_{i,0} 
    \end{array} \right),
\end{align}
while its antisymmetric counter part is
\begin{align}
   \sum_l \frac{(-1)^l}{2} \Delta_S^l(i) 
    & = \frac{i}{\sqrt{2}^3} \left (c_{i,1}^T,c_{i,2}^T\right)\sigma_y \left(\begin{array}{cc}
     1 & 0 \\
    0 & -1
    \end{array}\right) \left (\begin{array}{c}
         c_{i+1,1}  \\
         c_{i+1,2} 
    \end{array} \right) \notag\\
    & = \frac{i}{\sqrt{2}^3} \left (c_{i,\pi}^T,c_{i,0}^T\right)\sigma_y \left(\begin{array}{cc}
     0 & 1 \\
    1 & 0
    \end{array}\right) \left (\begin{array}{c}
         c_{i+1,\pi}  \\
         c_{i+1,0} 
    \end{array} \right).
\end{align}

 We next use the continuum limit $c_{i,a,\sigma} \simeq R_{a,\sigma}(x) e^{i k_a x_i} +L_{a,\sigma}(x) e^{-i k_a x_i}$, where $R,L$ are continuum fields of right and left movers. We omit any pair density wave, which implies that the antisymmetric component of the leg-pairing operator is dropped. Using $(-1)^a = 1$ ($(-1)^a = -1$) for $a = 0$, ($a = \pi$) we find
\begin{align}
\Delta_S^r(i) & \simeq \frac{-i }{\sqrt{2}} \sum_a (-1)^a R_{a,\sigma}(x_i) \sigma_y L_{a, \sigma}(x_i) \\
\Delta_S^l(i) & \simeq \frac{-i }{\sqrt{2}} \sum_a  \cos(k_a) R_{a,\sigma}(x_i) \sigma_y L_{a, \sigma}(i) \, .
\end{align}

\section{Further Numerical Results}
\label{app:further_results}

\subsection{Fourier transform of the pairing correlation functions}

\begin{figure}[h!]
    \centering
    \includegraphics{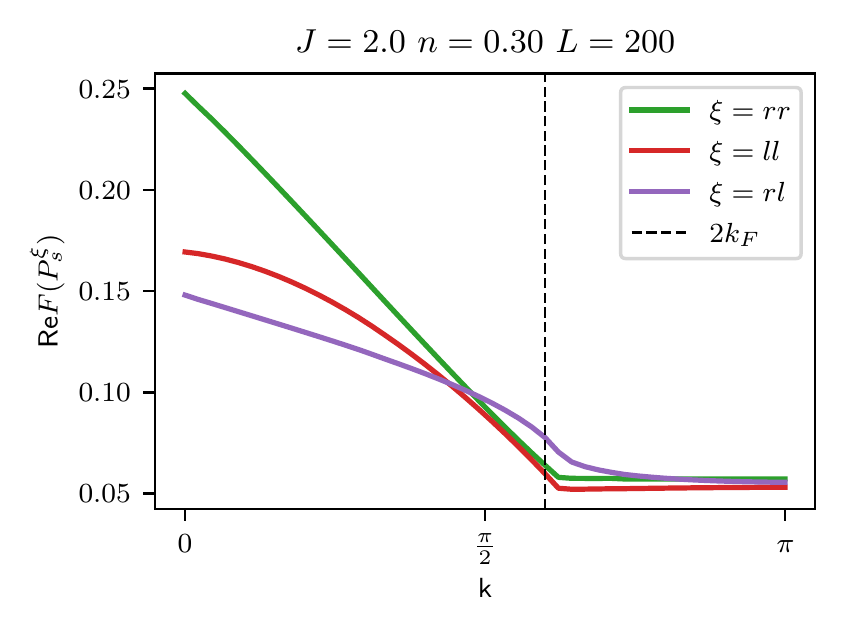}
    \caption{Real part of the Fourier transformation of the rung-rung, rung-leg and leg-leg pairing correlation functions for $J/t=2.0$ and $n=0.30$. In contrast to the rung-leg, the rung-rung and leg-leg correlation functions show a clear kink at $2k_F$.}
    \label{fig:fourier_pairing_correlations}
\end{figure}

Fig.~\ref{fig:fourier_pairing_correlations} displays the Fourier transform of the three pairing correlation functions treated by us.
As can be seen, the rung-leg correlator shows a smooth behavior around $2 k_F$, while the other two correlators show a kink in the Fourier transform, indicating oscillations in real space with wave number $2 k_F$. 
This indicates that further effects influence the behavior of the rung-leg correlator in real space, which make it difficult to use the ansatz Eq.~\eqref{eq:fit_correlation_functions} for obtaining meaningful values of the exponents for this quantity.

\subsection{Compressibility, spin gap, and charge gap for $V=0$}

\begin{figure}[h!]
		\centering 
		\includegraphics[]{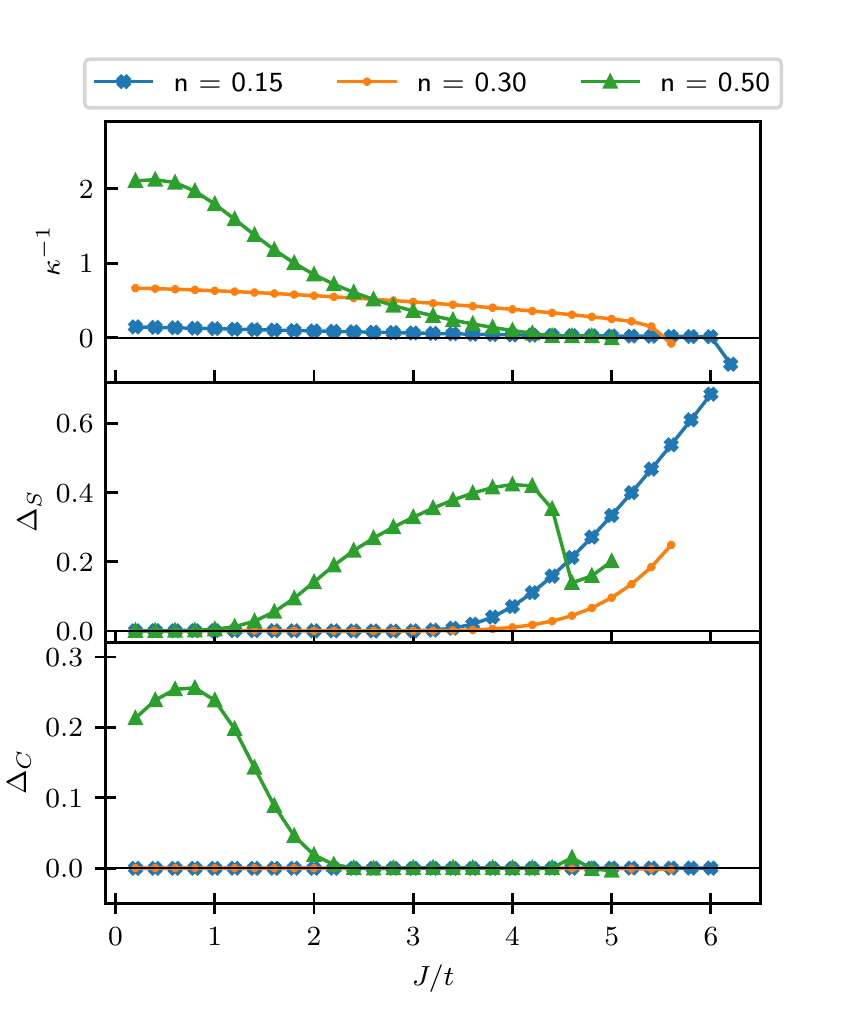}
		\caption{Comparison of the inverse compressibility (top), the spin gap (middle) and the two particle charge gap (bottom) for the $V=0$ \tJ ladder in the TL as function of $J/t$ for the densities $n=0.15$ (blue), $n=0.30$ (orange) and $n=0.50$ (green).}
	\label{fig:dSinverseKdC_V=0}
	\end{figure}

Fig.~\ref{fig:dSinverseKdC_V=0} shows the same quantities as Fig.~\ref{fig:dSinverseKdC_V=1}, but for the case $V=0$.
As can be seen, the overall behavior is similar, but the values of the phase transition points differ between both models.

\subsection{Charge structure factor for different system sizes}

	\begin{figure}
		\centering 
		\includegraphics{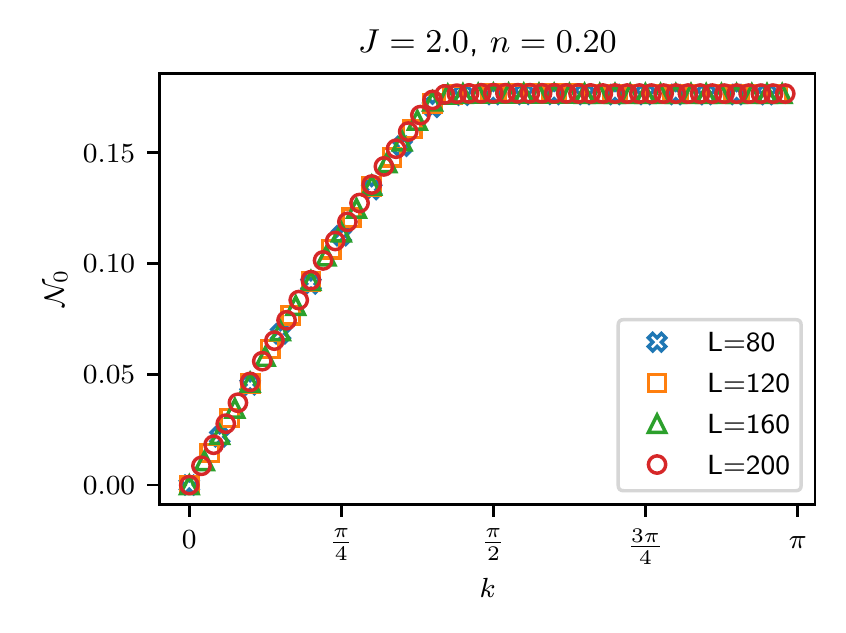}
		\caption{Comparison between the structure factor of the symmetric density correlation function for different system sizes for the regular two-leg \tJ ladder. Note that for $L<200$ the SymMPS energy convergence was set to $10^{-8}$.}
	\label{fig:compare_struct_factor}
	\end{figure}

Fig.~\ref{fig:compare_struct_factor} shows a comparison of the results for the symmetric part of the density structure factor for system sizes $L=80, \, \ldots, \, 200$.
As can be seen, finite size effects are not playing an important role for the case displayed.
Therefore, we focus on the behavior of the structure factor for the largest system sizes treated by us.

\begin{figure}[b]
	\includegraphics[width=0.45\textwidth]{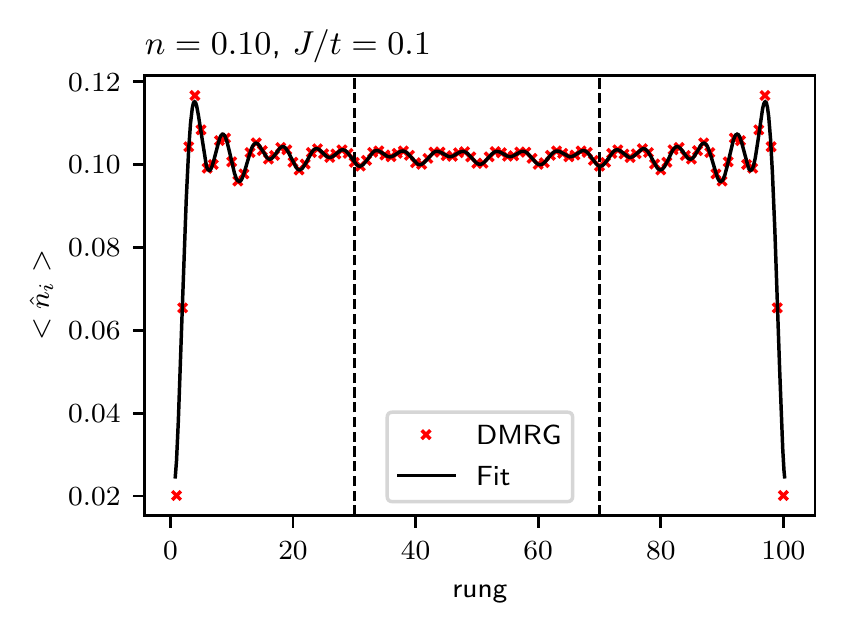}
	\caption{Expectation value of local electron density for $n=0.10$ and $J/t=0.1$ on one leg as function of the rung position. The black line denotes a fit applied to the data enclosed within the two dashed lines.}
	\label{fig:friedel_oscillations_fit}
\end{figure}

\subsection{Precursor Wigner crystal}

\begin{figure}[b]
	\centering
	\includegraphics[width=0.45\textwidth]{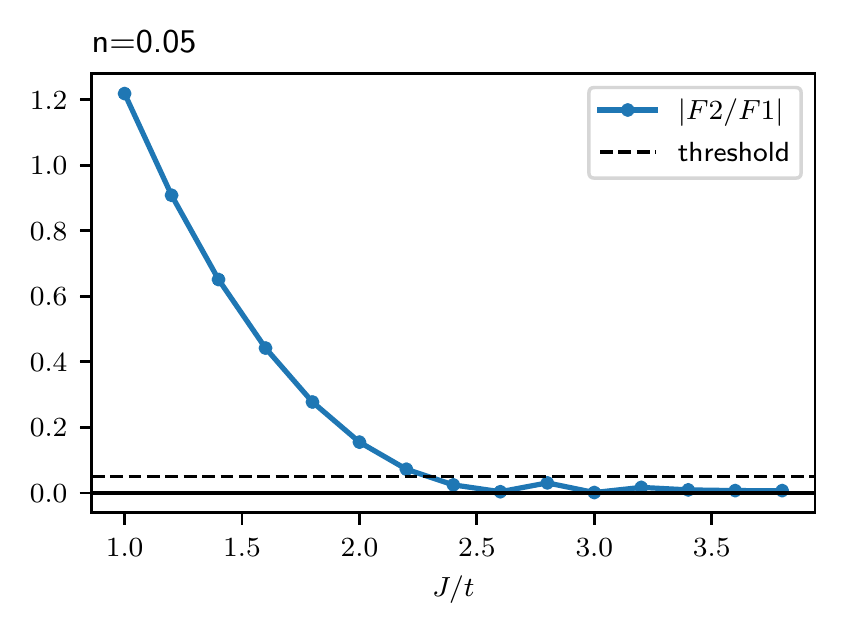}
	\caption{Ratio between the amplitudes, via fits of the local electron density, of the $2k_F$ (F1) and $4k_F$ (F2) term in Eq.~\eqref{eq:FriedelEggert} as function of $J/t$. The dashed line denotes a threshold where the $4k_F$ term becomes negligible. 
	}
	\label{fig:wigner_crystal}
\end{figure}

Figure~\ref{fig:friedel_oscillations_fit} shows an exemplary fit of the local electron density according to Eq.~\eqref{eq:FriedelEggert} in the Wigner crystal regime at $n=0.10$ and $J/t=1$ where $4k^0_F$ oscillations play an important role. 
Fig.~\ref{fig:wigner_crystal} shows the ratio $F_1/F_2$ i.e. the dominance of the $4k_F^0$ oscillations for $n=0.05$ as a function of $J/t$.

\section{Some details on the determination of the two-particle bound states at very low densities\label{app:two_el_solution}}

The general calculation idea was outlined by Lin~\cite{Lin1991}.
It makes use of a wave function ansatz
\begin{equation}
 \ket{\Psi} = \sum_{i_1, i_2} \Phi(i_1, i_2) c_{i_1\uparrow}^\dagger c_{i_2\downarrow}^\dagger \ket{0}
\end{equation}
and the explicit solution of the stationary Schrödinger equation (note the missing brackets in Eqs. (9) to (11) in \cite{Lin1991}!)
The energy of the two-particle bound state is determined by an implicit equation of the form
\begin{equation}
 \frac{r t^2}{J} = \frac{1}{I_0(E)} - E ,
 \label{eq:I_0_implicit}
\end{equation}
where $r = 8$ for a chain and $r = 16$ for a square lattice.
The integral $I_0(E)$ is defined (in the s-wave case for total momentum $\vec Q = 0$) as follows,
\begin{equation}
 I_0(E) = \frac{1}{2\pi} \int_{-\pi}^\pi \frac{\text{d}\vec q}{E - 2\epsilon_{\vec q}} \,.
\end{equation}
Although in our case we cannot assume the $90^\circ$ rotation symmetry, we can make a connection to the original calculation by taking the parity eigenbasis in the $y$-direction (corresponding to momenta $0$ and $\pi$) and halving of the respective matrix element $t$.
This yields
\begin{align}\begin{split}
  I_0(E) = \frac{1}{2\pi} \int_{-\pi}^\pi \text{d} q_x \; \Big( &\frac{1}{E + 2t + 4t\cos(q_x)} \\
  + &\frac{1}{E - 2t + 4t\cos(q_x)} \Big)
\end{split}\end{align}
and consequently Eq.~\eqref{eq:I_0_ladder}.
Furthermore, we find that in the case of the ladder $r = 12$ in \eqref{eq:I_0_implicit}.

\end{document}